\documentclass[aps,prl,amsmath, amssymb, english, aps, prl, twocolumn,reprint,floatfix, superscriptaddress]{revtex4-1}

\usepackage[english]{babel}
\usepackage{graphicx}
\usepackage{verbatim}

\usepackage{epsfig}
\usepackage{epstopdf}
\usepackage{amsfonts}
\usepackage{amsthm}
\usepackage{amsmath}
\usepackage{amssymb}
\usepackage{hyperref} 
\hypersetup{colorlinks=true}
\usepackage{makeidx}
\usepackage{pifont}
\usepackage{pdfpages}

\usepackage{color}
\usepackage{grffile}


\def\expect#1{\mathinner{\langle{#1}\rangle}}

{\catcode`\|=\active
  \gdef\expect#1{\left<\mathcode`\|"8000\let|\bravert {#1}\right>}}
\def\bravert{\egroup\,\vrule\,\bgroup}

\def\beq{\begin{equation}}
\def\eeq{\end{equation}}
\def\be{\begin{equation}}
\def\ee{\end{equation}}

\def\cG0{{\cal G}_0}

\def\vk{{\bf k}}

\def\a{\alpha}
\def\b{\beta}
\def\d{\delta}

\def\eps{\epsilon}

\def\g{\gamma}

\def\s{\sigma}
\def\uc2{$U_{c2}$}
\def\uc1{$U_{c1}$}

\def\bea{\begin{eqnarray}}
\def\eea{\end{eqnarray}}
\def \bal{\begin{align}}
\def \eal{\end{align}} %%aliases don't work in these environments, see ams FAQ
\def\#{\!\!}
\def\@{\!\!\!\!}

\def\+{\dagger}

\begin{document}

\title{\bf Selective Mottness as a key to iron superconductors}
\author{Luca de' Medici}
\address{Laboratoire de Physique et Etude des Mat\'eriaux, UMR8213 CNRS/ESPCI/UPMC, Paris, France}
\address{Laboratoire de Physique des Solides, UMR8502 CNRS-Universit\'e Paris-Sud, Orsay, France}
\author{Gianluca Giovannetti}
\address{CNR-IOM-Democritos National Simulation Centre and International School for Advanced Studies (SISSA), Via Bonomea 265, I-34136, Trieste, Italy}
%\affiliation{Institute for Theoretical Solid State Physics, IFW-Dresden, PF 270116, 01171 Dresden, Germany}
\author{Massimo Capone}
\address{CNR-IOM-Democritos National Simulation Centre and International School for Advanced Studies (SISSA), Via Bonomea 265, I-34136, Trieste, Italy}
\maketitle

{\bf 

The phase diagram of the high-T$_c$ cuprates is dominated by the Mott insulating phase of the parent compounds. As we approach it from large doping, a standard Fermi-liquid gradually turns into a bad non-Fermi liquid metal, a process which culminates in the pseudogap regime, in which the antinodal region in momentum space acquires a gap before reaching a fully gapped Mott state. 

Here we show that experiments for electron- and hole-doped BaFe${}_2$As${}_2$ support an analogous scenario. The doping evolution is dominated by the influence of a Mott insulator that would be realized for half-filled conduction bands, while the stoichiometric compound does not play a special role. Weakly and strongly correlated conduction electrons coexist in much of the phase diagram, a differentiation which increases with hole doping. 
We identify the reason for this selective Mottness in a strong Hund's coupling, which decouples the different orbitals. Each orbital then behaves as a single band Hubbard model, where the correlation degree only depends on how doped is each orbital from half-filling. 
Our scenario reconciles contrasting evidences on the electronic correlation strength and establishes a deep connection with the cuprates.
}

High-Tc superconductivity is the most spectactular example of a wide palette of unconventional behaviors induced by electronic correlations. The latter are indeed identified as the origin of both the Mott insulating parent compound and of the superconducting pairing in the cuprates. 
In this light the question of the correlation strength in iron superconductors\cite{Kamihara_pnictides2}, the runner-up materials in terms of $T_c$ (up to 57K), is crucial and remains to be clarified. Despite a bad metallic conductivity and many other unconventional properties\cite{Mazin_IronBoost,Johnston_Review_FeSC}, experiments have hardly detected any general direct signature of strong local interactions such as a Hubbard band\cite{Yang_Devereaux_weakcorr} and have instead reported an overall fair agreement with density-functional theory (DFT) predictions and assessed the local screened Coulomb repulsion U in the d-orbitals to be smaller than the total width of the conduction bands $W\sim4eV$.  This led to theories starting either from the weak-coupling\cite{Mazin_Splusminus} or from the strong-coupling\cite{Si_Abrahams_J1J2} pictures.

On the other hand, it has been emphasized early on that\cite{Haule_pnictides_NJP} that Hund's coupling J  plays a major role in determining the degree of correlations in these materials. This term, which favors high-spin configurations on each atom, is consistently estimated to be $\sim 0.3-0.6eV$ (in standard Kanamori notation) both theoretically and experimentally. On general grounds it has been found\cite{demedici_Janus} that for the filling of 6 electrons in 5 orbitals J favors a bad-metallic behavior yet pushing away the Mott insulator. It has also been shown\cite{Haule_pnictides_NJP} that this kind of correlated metal does not display as prominent Hubbard bands as in standard correlated materials. Overall an image of moderately strong correlations\cite{Johannes_local-moment} driven by Hund's coupling, seems to emerge.

The evolution as a function of doping is the smoking gun of the relevance of Mott physics (Mottness) inthe cuprates. Approaching half-filling the electronic motion is hindered leading to increasingly bad metals culminating in the pseudogap phase and eventually in an actual Mott insulator.
\begin{figure}[ht]
\begin{center} 
\includegraphics[width=8.5cm]{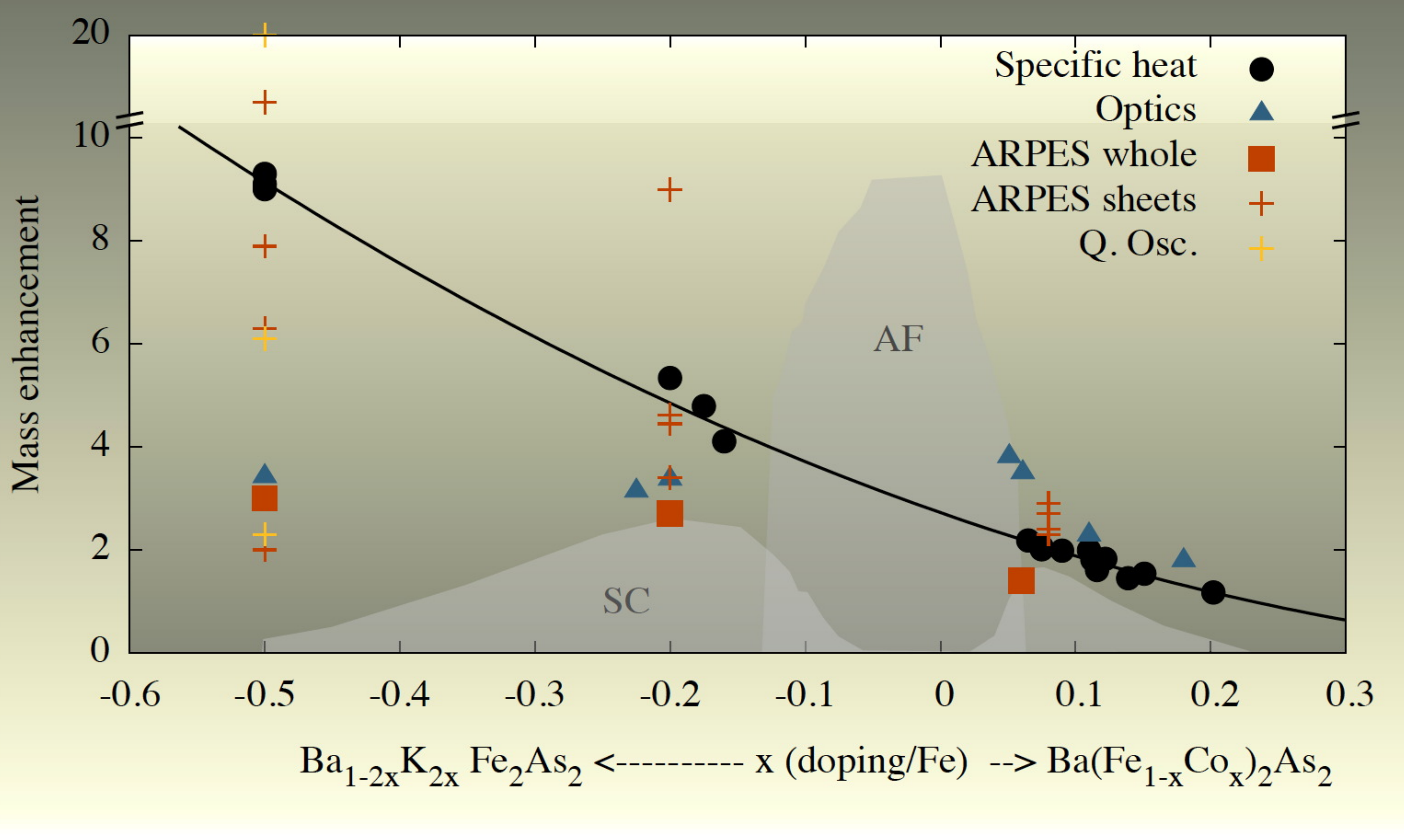}   
\caption{Experimental mass enhancements in doped BaFe${}_2$As${}_2$ assessed by different techniques (see online supplementary information for references). The experimental phase diagram, including the antiferromagnetically ordered metallic phase (AF) and the superconducting one (SC) is reproduced as a background. The solid line is a fit of the specific heat data. ARPES and quantum oscillation data corresponding to a given doping represent the values estimated for the different sheets of the Fermi surface. Correlations increase monotonously reducing doping, and the estimates from different techniques spread more and more (see text).}\label{fig:mass_enhancements}
\end{center}
\end{figure}

In the present paper we use the doping dependence as a litmus paper to assess the role of correlations in iron superconductors. By collecting experimental data from different probes, we show that in the 122 family (i.e. both hole- and electron-doped BaFe${}_2$As${}_2$) the effective mass of the carriers increases as the filling decreases and the stoichiometric compound is not special in terms of correlation strength. This doping dependence is heavily differentiated among the electrons forming the conductions bands. 

Our theoretical calculations will indeed show that the degree of correlation is determined by the distance to a Mott insulating state that would be realized with five electrons per iron atom. This effect is due to the dominance of the Hund's J, which in turn {\it decouples} the different orbitals, so that the multiorbital bandstructure becomes effectively a collection of single-band Hubbard models. Each of them happens to have a different population, which reflects in a different distance from the Mott state and in a different effective mass. This brings about a scenario of coexisting strongly and weakly correlated electrons. Indeed it was shown by two of us that iron superconductors are the ideal ground for heavily orbital-dependent correlation effects (selective Mottness), or even for orbital-selective Mott localization\cite{demedici_3bandOSMT, demedici_Genesis}. Phenomenological descriptions assuming the presence of itinerant and localized electrons\cite{Kou_OSMT_pnictides,Hackl_Vojta_OSMT_pnictides,Yin_Weiguo-Spin_fermion} have indeed reproduced many features of these materials.

Fig.\ref{fig:mass_enhancements} displays experimental estimates of the mass enhancement from specific heat, infrared reflectivity, angle-resolved photoemission spectroscopy (ARPES) and quantum oscillations (the references are detailed in the supplementary online information).  All the data are taken in the high-temperature tetragonal metallic phase above the magnetically ordered (and orthorombically distorted) phase at low dopings and the superconducting one at intermediate dopings. 
%We immediately notice  that while in the electron-doped side of the phase diagram the values are consistently grouped around $m^*/m_b \sim 2 \div 3$, moving through the stoichiometric filling towards the hole-doped side the estimated values spread more and more.

Single-crystal specific heat measurements delineate a very clear trend: the mass enhancement $m^*/m_b=\gamma/\gamma_b$ (where $\gamma$ and $\gamma_b$ are respectively the measured normal-state Sommerfeld coefficient and the band theory estimate $\gamma_b=\pi^2k_B^2/3N_0$,  $N_0$ being the total density of states at the Fermi energy) grows continuously from  $\sim 2$ in the electron-overdoped Ba(Fe${}_{1-x}$Co${}_x$)${}_2$As${}_2$ to 3 in the electron-underdoped/stoichiometric compounds towards 5-6 in the optimally hole-doped Ba${}_{1-x}$K${}_x$Fe${}_2$As${}_2$ until reaching the very high value of $\sim 9$ for the fully substituted KFe${}_2$As${}_2$ (having a doping of 0.5 holes/Fe).

Optical conductivity provides a partially different picture. Comparing the band-theory estimated of the Drude weight
\be\label{eq:Drude}
D_{band}=\sum_{\a,k} (\frac{\partial \epsilon^\a_k}{\partial k_x})^2 \delta(\epsilon^\a_k-\eps_F),
\ee
(where $\epsilon^\a_k$ is the dispersion of band $\a$) with the experimental value one obtains an optical effective mass $D_{band}/D_{exp}=m^*/m_b$.
With this approach  Qazilbash et al.\cite{Qazilbash_correlations_pnictides} found intermediate correlation strength ($\sim3.3$  for BaFe${}_2$As${}_2$ while the data we collect show correlations diminishing with electron doping but a possible saturation (with $m^*/m_b$ hardly exceeding $\sim 4$) of the mass enhancement for the hole-doped compounds. 

%We suggest a way to reconcile these seemingly contrasting evidences: we interpret these results as showing the coexistence of electrons with different degrees of renormalization. 

A general argument links this apparent discrepancy to the coexistence of electrons with different effective mass renormalization: in this multiband environment both the conductivity and the density of states at the Fermi energy stem from a sum over the band (or orbital) index. As a consequence $\gamma/\gamma_b=D_{band}/D_{exp}$ only if the mass enhancement is the same for all electrons. If instead the renormalization is different from band to band (from orbital to orbital) the density of states is a sum of contributions $\sim (m^*/m_b)_\a$ whereas the Drude weight is a sum of terms $\sim (m_b/m^*)_\a$, and thus as it happens for series or parallels of resistances, the first will be dominated by the largest mass enhancement, the second by the smallest.
% (albeit still weighted by the relative importance of these contributions). 
This means that in case of heavy differentiation of correlation strength the enhancement of the Sommerfeld coefficient will reflect that of the strongly correlated electrons, whereas the Drude weight the one of the weakly correlated ones.

A more direct confirmation of this picture is obtained by means of ARPES. In order to roughly match the experimental data the DFT bandstructure has to be renormalized by a global factor (indicated as `ARPES whole' in Fig. \ref{fig:mass_enhancements}) that increases with decreasing filling. Moreover by measuring the Fermi velocity on the different Fermi sheets and comparing them to DFT calculations one can estimate independently the mass enhancements of the corresponding low energy electrons. As we report in Fig.\ref{fig:mass_enhancements}, in the electron-doped side of the phase diagram the different Fermi sheets show a homogeneous renormalization of the Fermi velocities by a factor of order $\sim 2$, whereas moving to the hole-doped side the different Fermi sheets show different renormalization for different bands (between 3 and 9 for  Ba${}_{0.6}$K${}_{0.4}$Fe${}_2$As${}_2$  and between 2 and 18 for KFe${}_2$As${}_2$). Quantum oscillations measurements performed on KFe${}_2$As${}_2$ also report a heavy differentiation between the mass enhancement.

Thus in summary the unified phase diagram of electron- and hole-doped BaFe${}_2$As${}_2$ shows a strong and monotonous increase of correlations going from the heavy electron-overdoping to the heavy hole-overdoping, and a parallel progressive differentiation of the mass enhancements, or selective Mottness.
It is also worth noting that the stoichiometric filling does not play a particular role in this phase diagram of the tetragonal phase, contrary to the fact that commensuration is naively expected to strengthen the effects of correlations.
\begin{figure}
\begin{center} 
  \includegraphics[width=7cm]{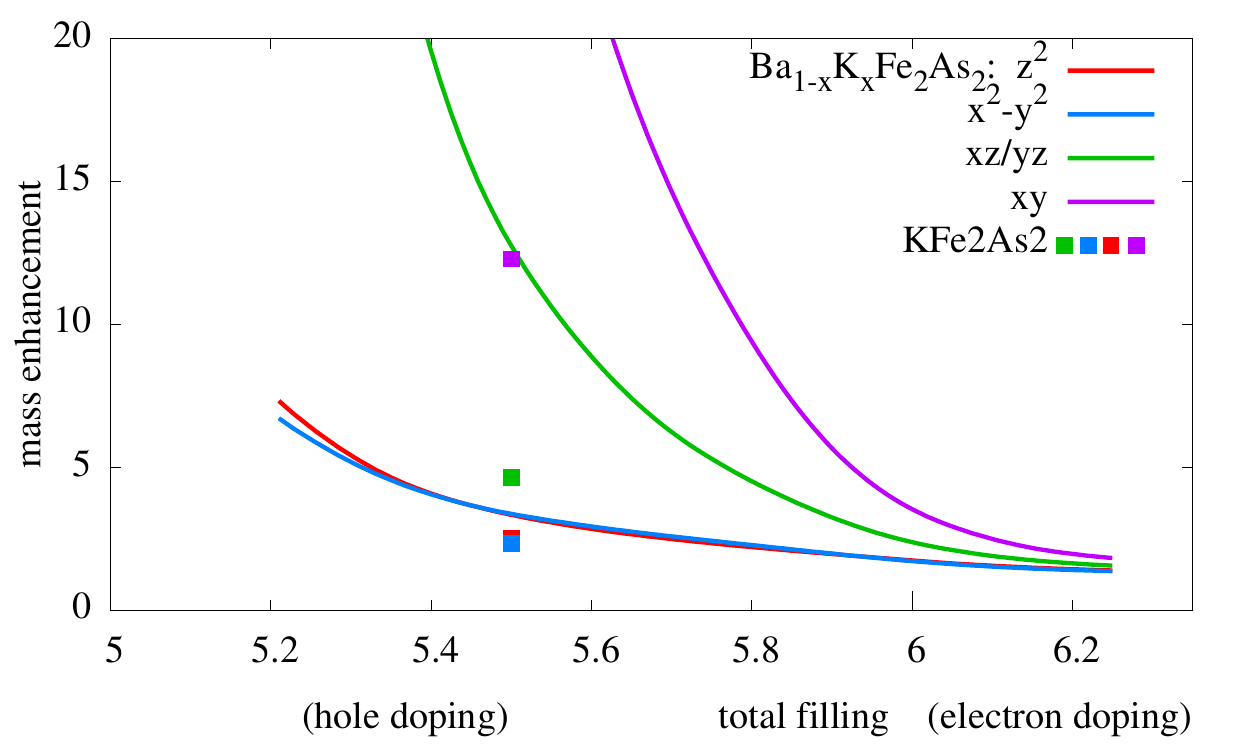}   
  \includegraphics[width=7cm]{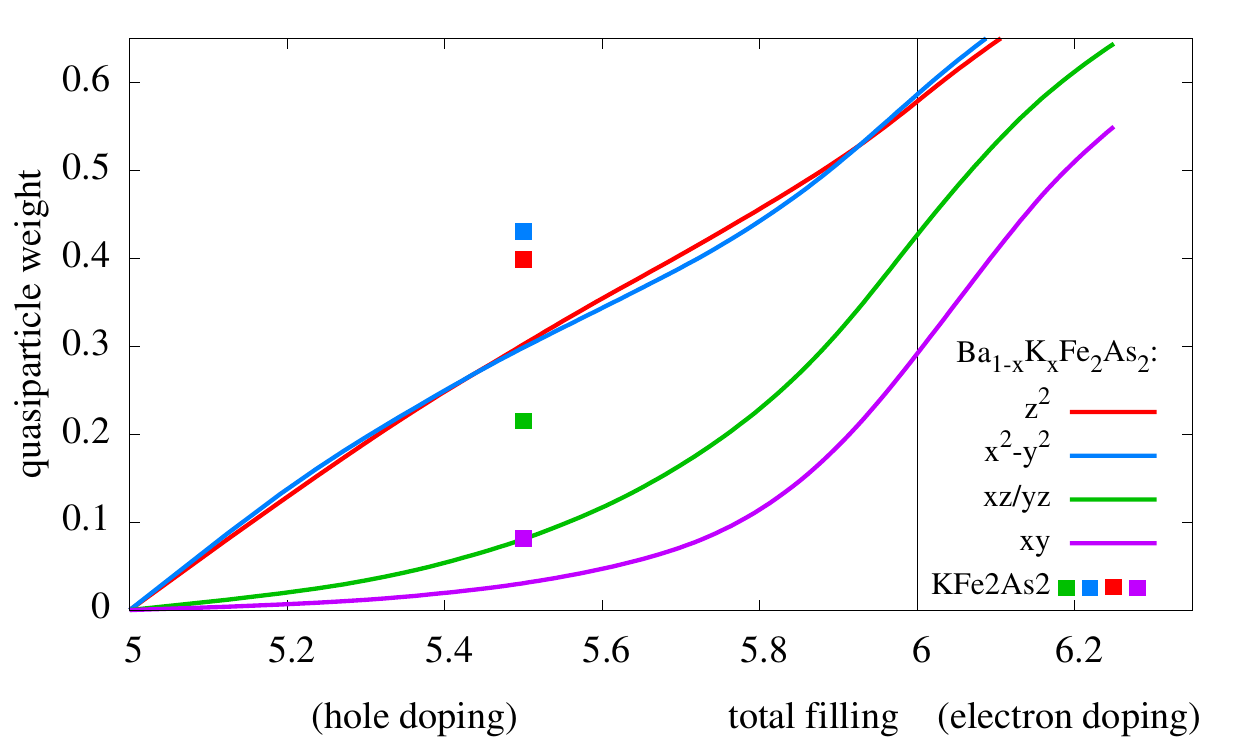}   
  \includegraphics[width=7cm]{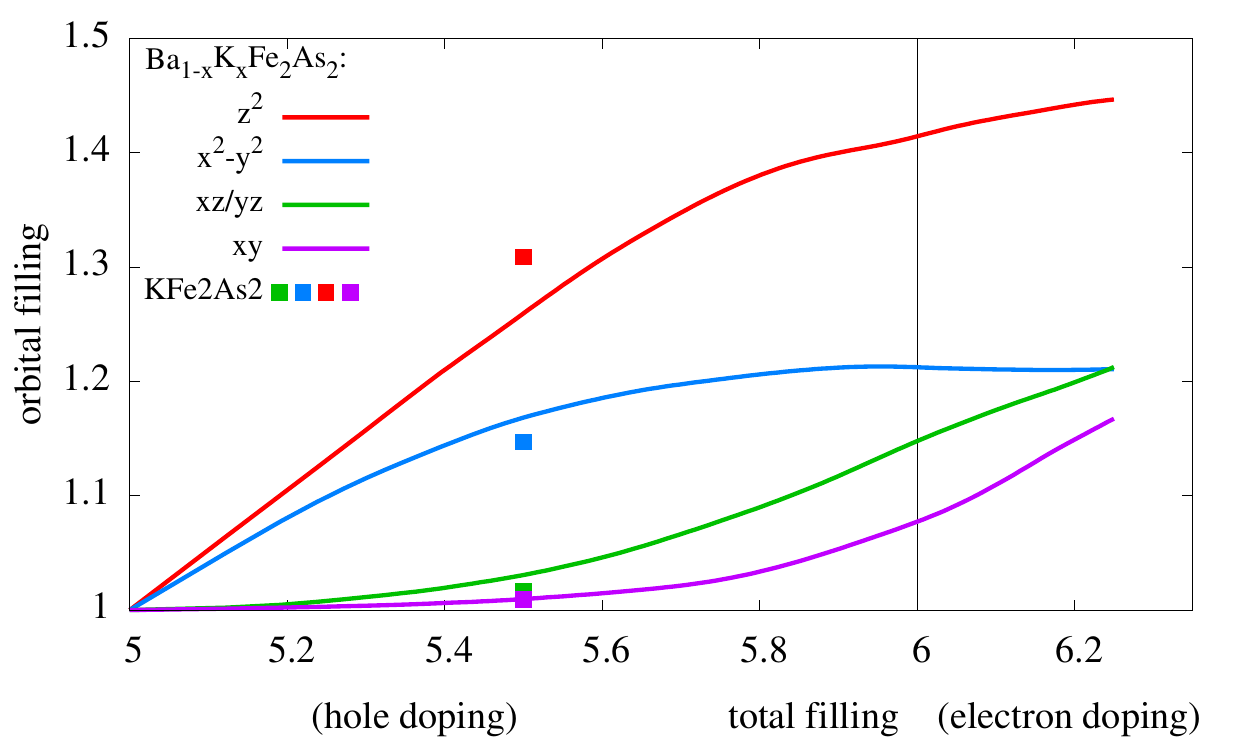}
   \includegraphics[width=7cm]{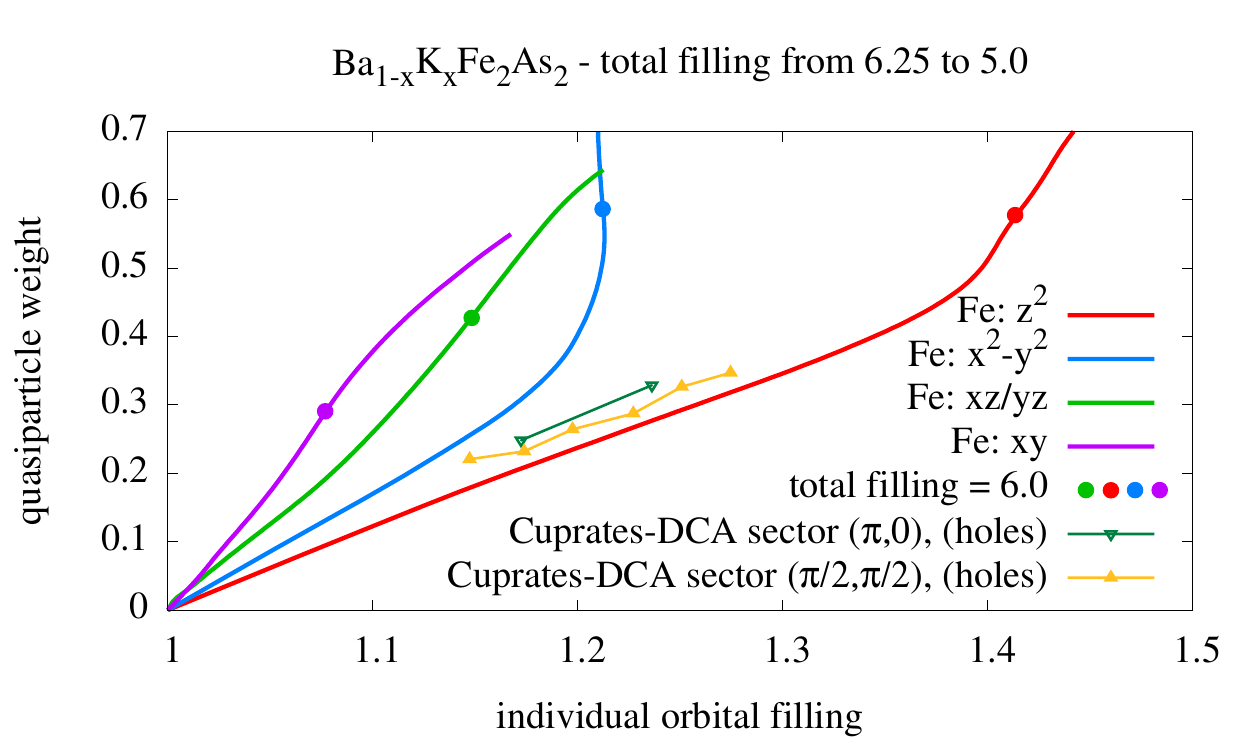}  
  \caption{Calculated orbitally resolved mass enhancement $(m^*/m_b)_\a$ and quasiparticle weight $Z_\a$ (two uppermost panels) and electron population (middle panel) for doped BaFe${}_2$As${}_2$ and stoichiometric KFe${}_2$As${}_2$ (squares)  within Slave-Spins mean-field. Note that for each orbital $\a$, $Z_\a$ tends to vanish when $n_\a$ approaches half-filling. Indeed when plotting $Z_\a$ as a function of $n_\a$ for each orbital (bottom panel) a striking linear behavior (typical of a doped single-band Mott insulator) appears showing that each orbital is only sensitive to its doping away from the half-filled Mott insulator found at total population n=5, independently from the others. The dots signal the values for the stoichiometric compound, having total population n=6. The same analysis is performed for DCA calculations on the bidimensional Hubbard model of Gull et al.\cite{Gull_DCA_k-space_selec}, representative of the pseudogap phase of underdoped cuprates,  pointing to a common mechanism.}
  \label{fig:Z_n_vs_doping}
\end{center}
\end{figure}

Calculations including dynamical correlations in DFT ab-initio calculations for both iron pnictides and chalcogenides show that the combined role of U and J gives rise to a novel scenario (as summarized in Ref. \cite{Georges_Annrev}). A sharp crossover\cite{Ishida_Mott_d5_nFL_Fe-SC, YuSi_LDA-SlaveSpins_LaFeAsO, Lanata_FeSe_LDA+Gutz} is found, as a function of increasing correlation strength and decreasing filling, towards a `Hund's metal'\cite{Yin_kinetic_frustration_allFeSC} state. The physics there is dominated by Hund's coupling and can be described using a few concepts isolated in model studies: 1) a strongly correlated metallic phase far from the Mott transition at n=6 \cite{demedici_Janus} is realized\cite{Aichhorn_Wannier_LaFeAsO,YuSi_LDA-SlaveSpins_LaFeAsO,Lanata_FeSe_LDA+Gutz} with a very low coherence temperature and correlation strength more sensitive to J than to U\cite{Haule_pnictides_NJP}; 2) above such coherence temperature a  `spin-frozen' state appears\cite{Liebsch_FeSe_spinfreezing,Werner_dynU_122}, in which the self-energy has a peculiar $\sqrt{\omega}$ dependence at low $\omega$ and long spin correlation time\cite{Werner_spinfreezing}; 3) `orbital decoupling'\cite{demedici_MottHund} happens  i.e. differentiation of the correlation strength among the different orbitals, due to the fact the Hund's coupling suppresses interorbital correlations, rendering charge excitations in the various orbitals virtually independent from one another.

This last point is what we highlight here as the main reason for the experimental scenario we have detailed. 
%Basically, when Hund's orbital decoupling is effective the correlation strength of each orbital is tuned individually, and it depends mainly on the doping of the orbital away from half-filling. 
%In this light, it should be taken into account that theoretical investigations involving only d electrons can overestimate the effect we discuss. According to DFT calculations, when we move from BaFe${}_2$As${}_2$ to KFe${}_2$As${}_2$,  the nominal reduction of the population of iron orbitals is compensated by an increased covalency with As p-orbitals, which partially counterbalances the effect of doping.

We have performed LDA+Slave-spin mean-field (see `Methods' section) calculations at $T=0$  starting from DFT bandstructures for doped BaFe${}_2$As${}_2$ in the tetragonal phase (see the Supplementary Information for other iron superconductors). 
We have checked that  for the interaction strength corresponding to the ab-initio estimate this compound displays the signature of the 'Hund's metal' phase. Our results as a function of U (see supplementary information) show that BaFe${}_2$As${}_2$ lies around or just above a cross-over to a correlated phase signaled by a fast drop in the quasiparticle weight $Z$, an increase of the interorbital spin correlation functions and decrease in the inter-orbital charge correlation functions, as reported in \cite{YuSi_LDA-SlaveSpins_LaFeAsO,Lanata_FeSe_LDA+Gutz}.
As expected in this regime the correlation strengths are strongly differentiated among orbitals, with the $t_{2g}$ orbitals more correlated the $e_g$ ones; in particular the $xy$ orbital is the most correlated of all. 
This motivates the experimental observation from ARPES that at low energy (where bands are mainly of $t_{2g}$ character) the bandstructure is more renormalized than at high energy (where both  $t_{2g}$ and  $e_g$ characters are present).
This heavy-fermionic behavior might be turned into an incoherent phase at $T\neq 0$ and/or with a refined treatment of quantum fluctuations\cite{Liebsch_FeSe_spinfreezing,Werner_dynU_122}. 

We have studied the evolution as a function of total density of the orbitally resolved quasiparticle weights $Z_\a= (m_b/m^*)_\a$ and populations, which are shown in Fig. \ref{fig:Z_n_vs_doping}. Correlations clearly increase continuously decreasing the total population, resulting in a more correlated hole-doped side compared to the electron-doped one\cite{Ikeda_pnictides_FLEX,Ishida_Mott_d5_nFL_Fe-SC,Misawa_d5-proximity_magnetic}. The commensurate filling of n=6 does not play any special role in this undistorted phase. 

It can be noted that the theoretical mass enhancements reach values, for strong hole overdoping, much higher than experiments. This is explained by a progressive increase of covalency  in the Fe-As planes introduced by the Ba $\rightarrow$ K substitution: when this is taken into account (using the DFT bandstructure for KFe${}_2$As${}_2$ at n=5.5) correlation strengths close to the experimental values are obtained (red dots in Fig.\ref{fig:Z_n_vs_doping}).

When the suggested `orbital decoupling' mechanism is realized the system switches from a collective to an individual orbital behavior: the correlation strength in each orbital depends mainly on the doping of the orbital away from half-filling. 
We have tested this hypothesis here and found that it is fully verified: in the lower panel of Fig. \ref{fig:Z_n_vs_doping} we plot the quasiparticle weight of each orbital as a function of its population. Each orbital is only sensitive to its own filling and behaves like a doped single-band Mott insulator, with a $Z$ characteristically proportional to the doping. With decreasing  total population all orbitals move towards half-filling and the quasiparticle weights go to zero at n=5, where the system is then a Mott insulator, with vanishing orbital polarization. The remarkable thing is that, albeit this happens at a very different rate for each orbital (the xy orbital is almost half-filled already at n=6, and thus more strongly correlated, the $z^2$ is quite far from half-filling until very close to n=5, and thus weakly correlated for most of the fillings) the behavior when plotted against the individual population is quite universal (the different slopes being determined by the details of the bandstructure).

This scenario confirms what observed in previous theoretical studies\cite{Ishida_Mott_d5_nFL_Fe-SC,Misawa_d5-proximity_magnetic}, that the parent compounds for pnictides at n=6, are influenced by correlations due to the Mott insulator at n=5.
Here we provide the reason why : the orbitals are `decoupled' from one another thus rendering the individual population of each orbital crucial for its correlation strength. Accordingly, the seemingly far in population n=6 compound, has orbitals instead very close to the Mott insulating state.  The $t_{2g}$ orbitals and the $xy$ in particular are only a few percents of doping away from the selective Mott insulator. The coexistence of electrons whose different correlation degree descends from a different distance to the Mott insulator defines the selective Mottness and its role in the phase diagram.

\begin{figure}
\begin{center} 
  \includegraphics[width=8.5cm]{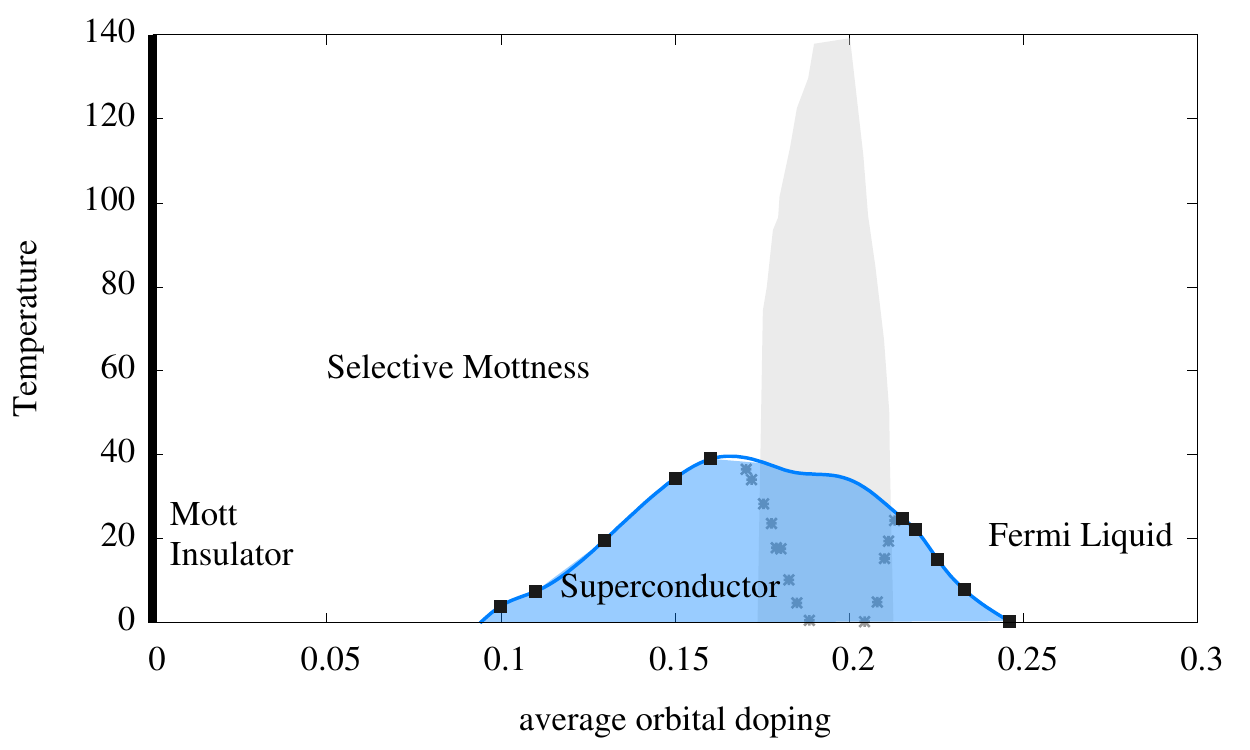}   
  \caption{Possible unified phase diagram for pnictides and cuprates. We plot here the experimental phase diagram for doped BaFe${}_2$As${}_2$ as a function of the average orbital doping. In the hypothesis that the magnetically ordered/orthorombically distorted phase (grey area) is accidentally favored by the total commensurate filling n=6 around the stoichiometric compound thus suppressing the superconductivity of the tetragonal phase, we artificially eliminate it and complete the superconducting dome (blue area). We then see that this dome is centered around 20\% doping (per orbital) away from the half-filled Mott insulator (here at n=5), as in cuprates. In between we find a strongly correlated Mott selective (and bad metallic \cite{Ishida_Mott_d5_nFL_Fe-SC}) phase, while at higher orbital dopings (corresponding respectively to the electron doped region in pnictides and the overdoped region in cuprates) we recover a moderately correlated Fermi liquid.}
\label{fig:provoc}
\end{center}
\end{figure}
One last more speculative hypothesis is now in order.
By plotting the experimental phase diagram as a function of the \emph{average orbital population} the stoichiometric filling falls at 20\% average doping\cite{Ishida_Mott_d5_nFL_Fe-SC}. This is now justified by the orbital-decoupled nature of the conduction electrons.

If, as quite generally believed, the magnetically ordered (and orthorombically distorted) phase is detrimental to superconductivity\cite{Mazin_IronBoost} and it is favored accidentally by nesting and/or lack of disorder realized for the commensurate n=6 filling, we can imagine that the unified phase diagram of the tetragonal phase of iron superconductors would see the two domes unified into a bigger one (see Fig. \ref{fig:provoc}).
This tentative phase diagram would now be strikingly similar to that of cuprate superconductors: a superconducting dome  would be centered around a doping of 20\% per band away from a Mott insulator, and flanked at smaller doping by  a bad metallic phase in which electrons with different correlation strength coexist, as it happens in hot/cold spots (culminating in the pseudogap phase) in the underdoped cuprates. At higher fillings a weakly correlated good metal is recovered in both cases. 
Indeed, as detailed in the supplementary information, the momentum-space selectivity describing the hot/cold spot and the pseudogap  phase of cuprates in the two-dimensional Hubbard model can be easily recast into a multi-orbital effective picture within cluster extensions of Dynamical Mean-Field methods\cite{Gull_DCA_k-space_selec}. 
The mass enhancement in each sector of the Brillouin zone scales linearly with the respective individual doping from half-filling (Fig. \ref{fig:Z_n_vs_doping}, lower panel), signaling that the orbital-decoupling mechanism is at play and reminding the situation outlined for iron superconductors.   
This scenario unifies a series of seemingly contradictory evidences and it is expected to put serious constraints on pairing theories for superconductivity.

\emph{Acknowledgments.} The authors acknowledge discussions with M. Aichhorn, R. Arita, E. Bascones, A. Georges, M. Imada, J. Mravlje, A. Millis, E.Winograd. 

\emph{Note} After this work was completed, a careful investigation of the coherence scales in KFe${}_2$As${}_2$ revealed heavy-fermionic behaviour\cite{Hardy_KFe2As2_Heavy_Fermion}, as predicted in our study.

L.d.M. is funded by CNRS and Agence Nationale de la Recherche under program ANR-09-RPDOC-019-01.
M.C. and G.G. are financed by ERC/FP7 through the Starting Independent Research Grant SUPERBAD (Grant Agreement 240524). Calculations have been performed on CINECA and CASPUR facilities.

\clearpage

%%%%%%%%%%%%%%%%%%%%%%%%%%%%%%%%%%%%%%%%%%%%%%%%%%%%%%%%
%%%%%%%%%%%%%%%%%%%%%%%%%%%%%%%%%%%%%%%%%%%%%%%%%%%%%%%%
%%%%%%%%%%%%%%%%%%%%%%%%%%%%%%%%%%%%%%%%%%%%%%%%%%%%%%%%
%%%%%%%%%%%%%  SUPPLEMENTARY MATERIAL %%%%%%%%%%%%%%%%%%%%%%%%%%

\begin{widetext}
\begin{center}
{\bf\Large Selective Mottness as a key to Iron superconductors\\ 
\bigskip
Supplementary Information}
\end{center}
\vspace{0.5cm}
\end{widetext}

\renewcommand{\thepage}{S\arabic{page}}  
\renewcommand{\thesection}{S\arabic{section}}   
\renewcommand{\thetable}{S\arabic{table}}   
\renewcommand{\thefigure}{S\arabic{figure}}

\setcounter{page}{1}
\setcounter{figure}{0}
\setcounter{table}{0}
%%%%%%%%%%%%%%%%%%%%%%%%%%%%%%%%%%%%%%%%%%%%%

\section{Experimental Estimates of the correlation}

We report here the values and references of the experimental data shown in Fig. 1 of the main text, where we plot several estimates of the quasiparticle mass enhancement in the tetragonal phase of the 122 pnictide family. We also outline a few caveats on the data analysis of the various experimental probes, and we specify the physical significance of the mass enhancement extracted with each method.
The original references are cited in the tables, and most of these results are nicely summarized in the reviews Refs. \cite{Stewart_RMP,SC_Johnston_Review_FeSC}.

\subsection{Specific Heat}

Let's first discuss the mass enhancement as estimated from the specific heat.
At low temperatures the specific heat of a Fermi-liquid goes like
\be
C_V=\gamma T +O(T^3)
\ee
where the ``Sommerfeld coefficient" $\gamma$ is proportional to the
renormalized mass of the carriers. The renormalization factor is then
usually estimated as $m^*/m=\gamma/\gamma_b$, where
$\gamma_b=\pi^2k_B^2/3N(\eps_F)$ and  $N(\eps_F)$ is the total
non-interacting band density of states at the Fermi energy $\eps_F$, as estimated by DFT.

The experimental values for $\gamma$ used to calculate the mass enhancements plotted in Fig.1 of the main text and  the corresponding DFT values, along with the original references are reported in table \ref{tab:SpecHeat}.

A source of error on $\gamma$ is however that in most of the cases reported here the specific heat of the normal tetragonal phase has to be extrapolated to zero temperature. Indeed the actual ground state is either a superconductor, or an orthogonally distorted phase with magnetic order, that we are not treating here.
What is done then in order to estimate $\gamma$ of the high-T tetragonal phase is to fit $C_V/T$ in the high-temperature phase and extrapolate it to zero T. 
\begin{table}[ht!]
\begin{tabular}{|c|rl|rl|c|}
\hline
\multicolumn{6}{|c|}{Specific heat - single crystals\hspace{0.5cm}{\scriptsize ($^1$ this work)}}\\
\hline
doping/Fe	&   $\gamma$   exp. &&	$\gamma_b$ DFT &&	$m^*/m=\gamma/\gamma_b$\\
\hline	
0.202 &   11.03&\cite{Hardy_SpecHeat_Co122}	&	9.44&$^1$  &	1.17 	\\
0.151  & 15.84&\cite{Hardy_SpecHeat_Co122}	&	10.32 &$^1$	 &1.53	\\
0.139  & 14.11&\cite{Hardy_SpecHeat_Co122}	&	9.75 &$^1$ 	&1.45	\\
0.122  & 17.22&\cite{Hardy_SpecHeat_Co122}	&	9.44 &$^1$ 	&1.82	\\
0.116  & 15.11&\cite{Hardy_SpecHeat_Co122}	&	9.41&$^1$  	&1.60	\\
0.1125&17.02&\cite{Hardy_SpecHeat_Co122} 	&	9.41&$^1$  	&1.80	\\
0.11    &	18.67&\cite{Hardy_SpecHeat_Co122}	&	9.38&$^1$  	&1.99 \\
0.09    &	20.06&\cite{Hardy_SpecHeat_Co122}	&	10.12&$^1$ 	&1.98\\
0.075  &	 22.02&\cite{Hardy_SpecHeat_Co122}	&	10.96&$^1$ 	&2.0	\\
0.075  &	22.53&\cite{Hardy_SpecHeat_Co122}	&	10.96&$^1$	&2.05	\\
0.065  &	24.06&\cite{Hardy_SpecHeat_Co122}	&	11.03&$^1$ 	&2.18	\\
-0.16   &	50&\cite{Popovich_SpecHeat_BaK0.32-122}		&	12.17&$^1$ 	& 4.09		\\
-0.175  &   57.5   &\cite{Pramanik_SpecHeat_BaK122}	&12.00& $^1$ 	& 4.66		\\
-0.20   &	63 &\cite{Mu_SpecHeat_BaK0.4-122}		&	11.80 &$^1$ 	&5.01			\\
-0.5	    &91&\cite{Kim_SpecHeat_KFe2As2}		&	10.1&\cite{Terashima_Qosc_KFe2As2}  	&9.0			\\
%-0.5	    &92	&Hashimoto	&	10.1&\cite{Terashima_Qosc_KFe2As2}  	&9.1		\\
-0.5	    &94&\cite{Abdel-Hafiez_SpecHeat_KFe2As2}		&	10.1&\cite{Terashima_Qosc_KFe2As2}  	&9.3			\\
\hline
\end{tabular}
\caption{Mass enhancements estimated from specific heat measurements used in Fig. 1 of the main text. The columns report respectively the values for the Sommerfeld coefficient $\gamma$ extracted from experiments, the corresponding DFT calculated values $\gamma_b$ and the calculated mass enhancement.  Several LDA values for $\g_b$ are available for KFe${}_2$As${}_2$ in the literature ranging from 10.1\cite{Abdel-Hafiez_SpecHeat_KFe2As2,Terashima_Qosc_KFe2As2} to 13.0\cite{Abdel-Hafiez_SpecHeat_KFe2As2}. Our calculation yields 14.5, thus confirming the wide error bar associated to this value. The estimated mass-enhancement ranges then form 6.5 to 9.3. We use the more dramatic value of 9.3 in the Fig. 1 of the main article in order to better illustrate this effect, but all values are higher than the ones estimated anywhere else in the phase diagram, thus confirming the solid physical claim of increasing correlations estimated with specific heat, with reducing total population.}
\label{tab:SpecHeat}
\end{table}
When the low-T phase is superconducting this fit can be helped by the
additional constraint due to the second-order nature of the
superconducting transition, that imposes matching entropies $S(T)=\int
dT C_V/T$ at $T_c$. Then from the actual measurement of $C_V$ in the
superconducting phase one can impose a constraint on the integral of
$C_V/T$ of the normal phase in the extrapolated range of temperatures.
No similar constraint can be used around the stochiometric $n=6$
compound, where the system undergoes a first-order magnetic/structural
transition at $\gtrsim100K$, thus making the extrapolation quite
unreliable. For this reason we omitted the data in this region. It is worth reminding that the reliability of these estimates are also subject to the ability to subtract the phononic contribution to the specific heat, that has to be estimated independently.

\begin{figure}
\begin{center}
\includegraphics[width=8cm]{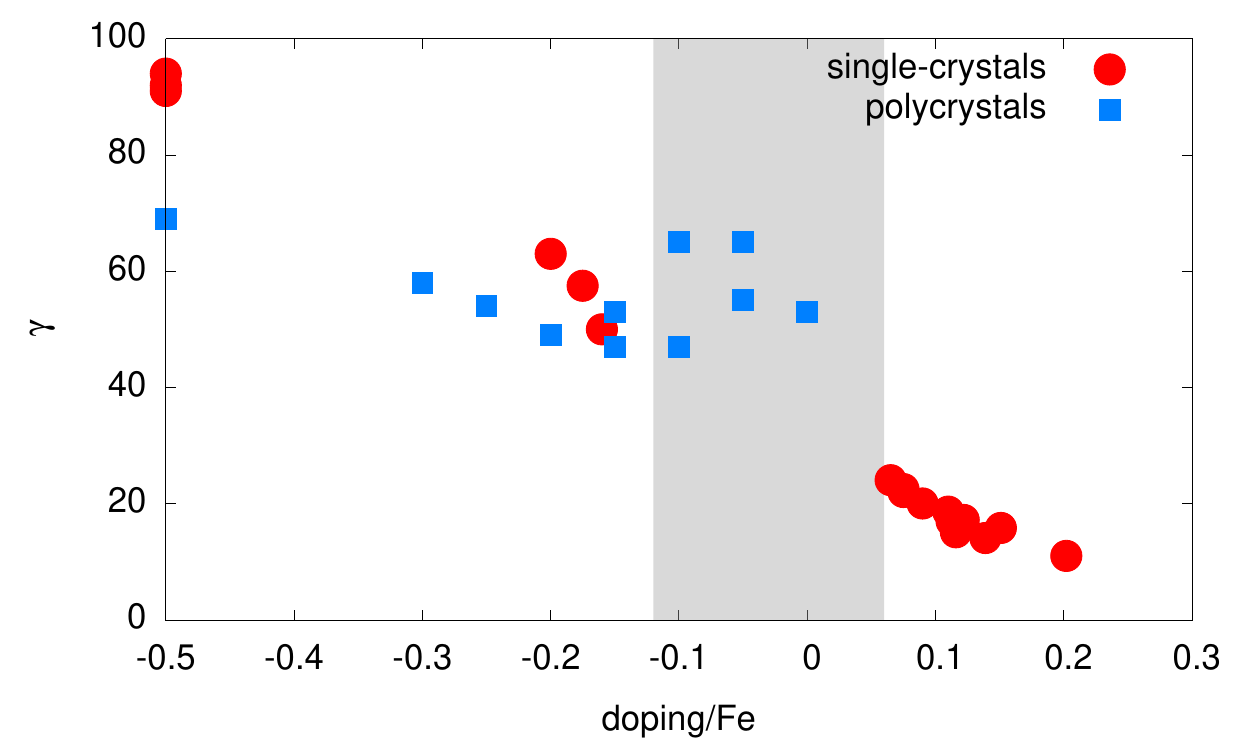}
\end{center}
\caption{Sommerfeld coefficient $\gamma$ for single-crystals used in Fig.1 of the main text, compared with the values measured in polycrystals. The gray area represents the zone where the low-T phase is orthorombical distorted and magnetically ordered, and where the specific heat of the high-T tetragonal paramagnetic phase has to be extrapolated from above the SDW transition. Outside this zone, where the error bars on $\gamma$ are necessarily higher, even polycrystals show an enhancement of the Sommerfeld coefficient with hole doping. Polycrystal data are taken from \cite{Storey_SpecHeat_BaK122_polycr, Kant_SpecHeat_BaK122_polycr,Fukuzawa_SpecHeat_KFe2As2}}\label{fig:Cfr_single_poly}
\end{figure}

Also, the purity of the sample is of utter importance.
Only measurements performed on single crystals are shown in Fig.1 of the main text, specific heat in polycrystals being subject to substantial variations. An example is the substantially reduced value $\gamma=69 mJ/mol K^2$ reported\cite{Fukuzawa_SpecHeat_KFe2As2} for polycrystalline KFe${}_2$As${}_2$ of residual resistivity ratio RRR=67, to be compared with $\gamma=94 mJ/mol K^2$  reported\cite{Kim_SpecHeat_KFe2As2} for a single crystal with RRR=650. See Ref. \cite{Stewart_RMP} for a somewhat more thorough discussion on this point. 
This being said, where their estimate is more reliable also polycrystals show an increasing Sommerfeld coefficient with hole doping (see Fig. \ref{fig:Cfr_single_poly}).

\subsection{Optical conductivity}

Other mass enhancements reported in Fig.1 of the main text are
extracted from optical conductivity.  Indeed the peak in the spectrum
at low frequency (the Drude contribution) is usually considered to
account for the quasiparticle contribution to the conductivity.
 Its width accounts for the scattering rate of these charge carriers,
 whereas its total weight for their effective mass. Comparing this
 weight with the value obtained in DFT is a possible way to assess the mass enhancement due to dynamical correlations. Indeed correlations reduce the Drude weight, and in a Mott insulator its value vanishes.

This can be seen as an analysis of the reduction of the kinetic energy of the quasiparticles\cite{SC_Qazilbash_correlations_pnictides} due to correlations,  $K_{exp}/K_{band}$, that indeed vanishes in a Mott insulator. The overall kinetic energy of the conduction electrons however, is reduced by correlations but does not vanish, even in a Mott insulator where itinerant (albeit incoherent) excited states still exist at higher energy. The spectral weight of intraband transitions always adds up to the conduction electrons kinetic energy\cite{Millis_optics_lectures} but in correlated materials it is transferred away from the Drude peak into mid-infrared (MIR) and higher energy features.
Thus an alternative, approximate way to estimate the relative
reduction of the quasiparticle kinetic energy based exclusively on
experimental data was proposed by Lucarelli et
al.\cite{Lucarelli_Optics_Co122,DeGiorgi_Correl_Optics_pnictides_beyond}. It
consists in normalizing the spectral weight of 
 the Drude peak to the Drude+MIR total spectral weight. The latter is indeed an underestimate of the total band-theory kinetic energy, but has the advantage of excluding inter-band transition that contaminate the spectral weight at still higher energies.

In Ref. \cite{DeGiorgi_Correl_Optics_pnictides_beyond} it was shown that mass enhancements estimated through this procedure agree rather well with the estimate $K_{exp}/K_{band}$ in BaFe${}_2$As${}_2$ of Ref. \cite{SC_Qazilbash_correlations_pnictides}. Thus this analysis was performed on the electron-doped compounds in \cite{Lucarelli_Optics_Co122} using two methods: the Drude and MIR contribution were determined using a Drude-Lorentz model and then integrated separately, or they were approximately separated using two appropriately chosen cutoffs.
The mass enhancement was defined as:
\be
 m^*/m\simeq K_{band}/K_{exp}\simeq SW_{Drude+MIR}/SW_{Drude},
\ee
with $SW(\Omega_c)=\int_0^{\Omega_c} d\omega \s^\prime(\omega)$, where $\s^\prime(\omega)$ is the real part of the optical conductivity, $\Omega_c=500 cm^{-1}$ for the Drude contribution and $\Omega_c=2000 cm^{-1}$ for Drude + MIR.

Here we extend this last method to available experimental data on the hole-doped side and report the obtained mass enhancements in table \ref{tab:optics}.
\begin{table}
\begin{tabular}{|l|l|l|}
\hline
\multicolumn{3}{|c|}{Optics}\\
\hline
doping/Fe	& $m^*/m$ (fit)  & $m^*/m$ (cutoffs)\\
\hline	
0.18	  	&	1.48\cite{Lucarelli_Optics_Co122}		&	1.79\cite{DeGiorgi_Correl_Optics_pnictides_beyond}\\
0.11  	&	1.98\cite{Lucarelli_Optics_Co122}		&	2.28\cite{DeGiorgi_Correl_Optics_pnictides_beyond}\\
0.061  	&  	4.21\cite{Lucarelli_Optics_Co122}		&	3.50\cite{DeGiorgi_Correl_Optics_pnictides_beyond}\\
0.051  	&  	4.56\cite{Lucarelli_Optics_Co122}		&      3.80\cite{DeGiorgi_Correl_Optics_pnictides_beyond}\\
0.025	&	4.78\cite{Lucarelli_Optics_Co122}		&	4.99\cite{DeGiorgi_Correl_Optics_pnictides_beyond}\\	
0.0	   	&	 3.3\cite{Lucarelli_Optics_Co122,SC_Qazilbash_correlations_pnictides}& 	4.80\cite{DeGiorgi_Correl_Optics_pnictides_beyond}\\
-0.2	  	&	 & 3.31\cite{Li_Optics_BaK-122}\\
-0.225   	&	 & 3.15\cite{Yang_Optics_BaK0.45_low_energy_bos}\\
-0.5	  	&	 &	3.37\cite{Wang_Optics_doped122}\\
\hline
\end{tabular}
\caption{Mass enhancement values extracted from optical conductivity measurements and used in Fig. 1 of the main text. The mass enhancement is estimated via the ratio of the Drude spectral weight and the Drude+MIR spectral weight. Two method are used to estimate these spectral weigths, as described in the text. The first (results in columns 2) uses a fit to Drude-Lorentz oscillators, the second (column 3, the hole-doped values are integrals calculated by digitizing the data reported in the cited references) uses two cutoffs: $\Omega_c=500 cm^{-1}$ for the Drude component, and $\Omega_c=2000 cm^{-1}$ for the Drude+MIR weight. The values of column 3 are used in Fig.1 of the main text.}
\label{tab:optics}
\end{table}

 It is worth pointing out that several technical points make this
 analysis highly nontrivial and are possible sources of error. It is
 difficult for instance to isolate the Drude contribution within the
 optical conductivity: indeed interband transitions partially
 superpose to it, especially when the band structure is as intricate
 as in iron superconductors, where multiple low-energy interband transitions
are available.
This  makes the choice of a cutoff for the spectral weight integral
uneasy. On the other hand fits based on the Drude-Lorentz model can be
used to sort out the Drude contribution, but often several
alternatives are possible. An ubiquitous complication is the need of
at least two Drude components in order to fit the low-energy spectral
weight. One Drude peak is narrow while the other is very wide,
accounting for incoherent processes and overlapping with interband
contributions. Last but not least, the coupling to bosonic modes (as magnetic fluctuations for example) can also shift some spectral weight, rendering the fits or the choice of the cutoffs even harder.
 
Overall then, the mass enhancements estimated through the analysis of the optical spectral weight that we have performed have to be considered as indicative. And indeed they yield a picture of intermediate correlation strengths throughout the phase diagram of doped BaFe${}_2$As${}_2$, turning to rather weaker correlations in the electron-overdoped region\cite{Drechsler_Optics_122_interm_corr}.

\subsection{ARPES and Quantum Oscillations}

\begin{table}
\begin{tabular}{|l|l|l|l|l|l|l|l|l|}
\hline
\multicolumn{9}{|c|}{ARPES}\\
\hline
doping &	whole &	\multicolumn{5}{|c|}{sheets} & \hspace{0.15cm} $z^2$ &  Ref.\\
\cline{3-7}
per Fe	    &			    & $\a$ & $\b$ &$\zeta$ &$\gamma$ &$\d$/$\eps$& band		&	\\
\hline
0.08 	&		    &	 2.7	& 2.3 & 	&  2.4	&	2.9   &		&\cite{Brouet_orb_res_lifetimes_Co122}\\
0.06 &	1.4	    & & & & & &&\cite{Yi_Shen_ARPES_hole_electron_doped122}\\
-0.2		&	2 (2.7\cite{Yi_Shen_ARPES_hole_electron_doped122})	    & 3.4	& 4.45	& 	& 4.62	& 	9.0	 & 		&\cite{Ding_Arpes_BaK}\\
-0.5	 	&	3	    &	 2.0	& 6.3	&  7.9 &	&	18.7	 &	3	&\cite{Yoshida_ARPES_KFe2As2}\\
\hline
\multicolumn{9}{|c|}{Quantum oscillations}\\
\hline
-0.5		&	& 2.3 & 	&  6.1 &  & 20 &	&\cite{Terashima_Qosc_KFe2As2}\\
\hline
\end{tabular}
\caption{Mass enhancement values extracted from ARPES and Quantum Oscillation measurements and used in Fig. 1 of the main text. What we highlight in this paper is that they spread moving from electron-doping to hole-doping. The first column reports the rough proportionality factor in order for the measured bandstructure and the DFT one to best match each other globally (showing that also the degree of correlation increases going from electron- to hole-doping. The following columns report the mass enhancements estimated as the ratio between the theoretical and measured Fermi velocities, resolved for every Fermi sheet. $\a$,$\b$ and $\zeta$ are hole-like pockets, whereas $\g$, $\d$ and $\eps$ are electron-like ones. The electron doped data are taken at $k_z\sim 1$ ((i.e. along $Z-A$) whereas the hole-doped are taken at $k_z\sim 0$ (i.e. along $\Gamma-M$. The last column reports the mass enhancement of the band of main character $3z^2-r^2$ that lies below the Fermi level and slightly crosses it near Z in KFe${}_2$As${}_2$. Finally Quantum Oscillations measurements (that do not observe the $\b$ Fermi sheet) for KFe${}_2$As${}_2$ are reported.}
\label{tab:ARPES_QOsc}
\end{table}

We report in table \ref{tab:ARPES_QOsc} the mass enhancements extracted from ARPES measurements in both electron- and hole-doped BaFe${}_2$As${}_2$ and in KFe${}_2$As${}_2$, for which we also report Quantum Oscillation measurements.

What we highlight is that going from electron-doping to hole-doping until reaching KFe${}_2$As${}_2$ (0.5 holes/Fe doping)  correlation strength increases (looking at the global renormalization factor needed in order to match the theoretical and the observed bandstructure). These data also show the measured mass enhancement for each Fermi sheet (each of which bears a different dominant orbital character), spread more and more with the reduction of the filling, thus confirming the increasing orbital selectivity.

A caveat is due however when directly comparing ARPES and quantum
oscillations with the DFT band-structures. Electronic correlations
induce both bandwidth renormalizations and band shifts. This can
result in a change of the location (and size) of the Fermi surface and
thus comparing the electron velocities at the Fermi level can be
misleading. In principle a safer procedure would be to compare the
measured Fermi velocity with the calculated velocity of bands once
shifted in order to better match
experiments\cite{Brouet_doping_dep_shifts}.
Despite the sizable error bars induced by this effect, the large spread of the mass renormalizations
depending on the different character of electrons at the Fermi level is a clear trend.

%is a physically solid feature, we believe (con la faccia sotto i suoi
%piedi e si puo' anche muovere).

\subsection{Other experimental data supporting Mott Selectivity in Iron Superconductors}

Besides some of the cited studies (the optics papers (Lucarelli et al.\cite{Lucarelli_Optics_Co122} and Wang et al.\cite{Wang_Optics_doped122}) interpreting results as a coexistence of more itinerant and more localized electrons and the ARPES studies by Ding et al.\cite{Ding_Arpes_BaK} and Yoshida et al.\cite{Yoshida_ARPES_KFe2As2} highlighting the strong orbital dependence of  correlations) other experiments point towards a main role played by Mott selectivity in Iron pnictides.
Yuan et al. \cite{Yuan_magnetoresistance_122_local-itinerant} argue that magnetotransport data in Co-doped BaFe${}_2$As${}_2$ can be interpreted as showing that both itinerant and localized electrons are present, and that the scattering of the former on the latter gives rise to the complex transport properties that are measured.
Malaeb et al.\cite{Malaeb_abrupt_gap_change_BaK122} have shown a strong orbital-dependence of the superconducting gap with hole-doping, leading to a disappearance of only one gap in superconductive Ba${}_{0.4}$K${}_{0.6}$Fe${}_2$As${}_2$

Also in the arguably more correlated Iron chalcogenides several experiments pointed towards a localized-itinerant dichotomy.
NMR and EPR measurements  on FeSe${}_{0.42}$Te${}_{0.58}$ by Ar\v{c}on et al. \cite{Arcon_NMR_EPR_FeSeTe_local-itinerant}, and neutron scattering measurement on FeSe${}_{0.65}$Te${}_{0.35}$ by Xu at al. \cite{Xu_neutrons_FeTeSe_local-itinerant} are intepreted as  showing the presence of intrinsic local magnetic moments in the metallic non-superconducting phase, coexisting then with itinerant electrons.
Also Tamai et al. \cite{Tamai_ARPES_FeSeTe_strongcorr} report strong orbital differentiation and in particular stronger correlations for the xy orbital in  FeSe${}_{0.42}$Te${}_{0.58}$, while Yi et al. \cite{Yi_Shen_ARPES_OSMT_KFeSe} study the intercalated chalcogenides A${}_x$Fe${}_{2-y}$Se${}_{2}$ (A=K, Rb) and report the disappearance of the band of xy-character when the temperature is increased above $\sim 150K$, thus signaling the orbital-selective Mott transition of the most correlated electrons.

\newpage

\begin{widetext}

\begin{figure}[ht]
\begin{center} 
  \includegraphics[width=5cm]{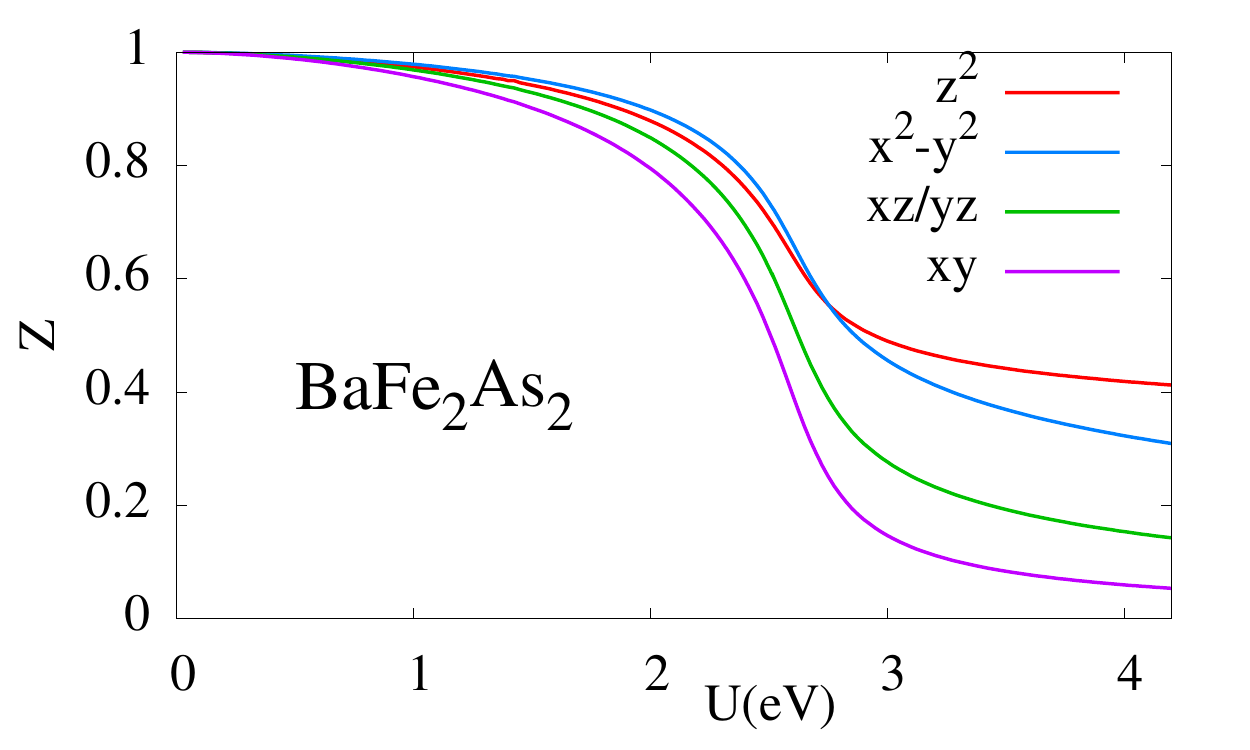}
  \includegraphics[width=5cm]{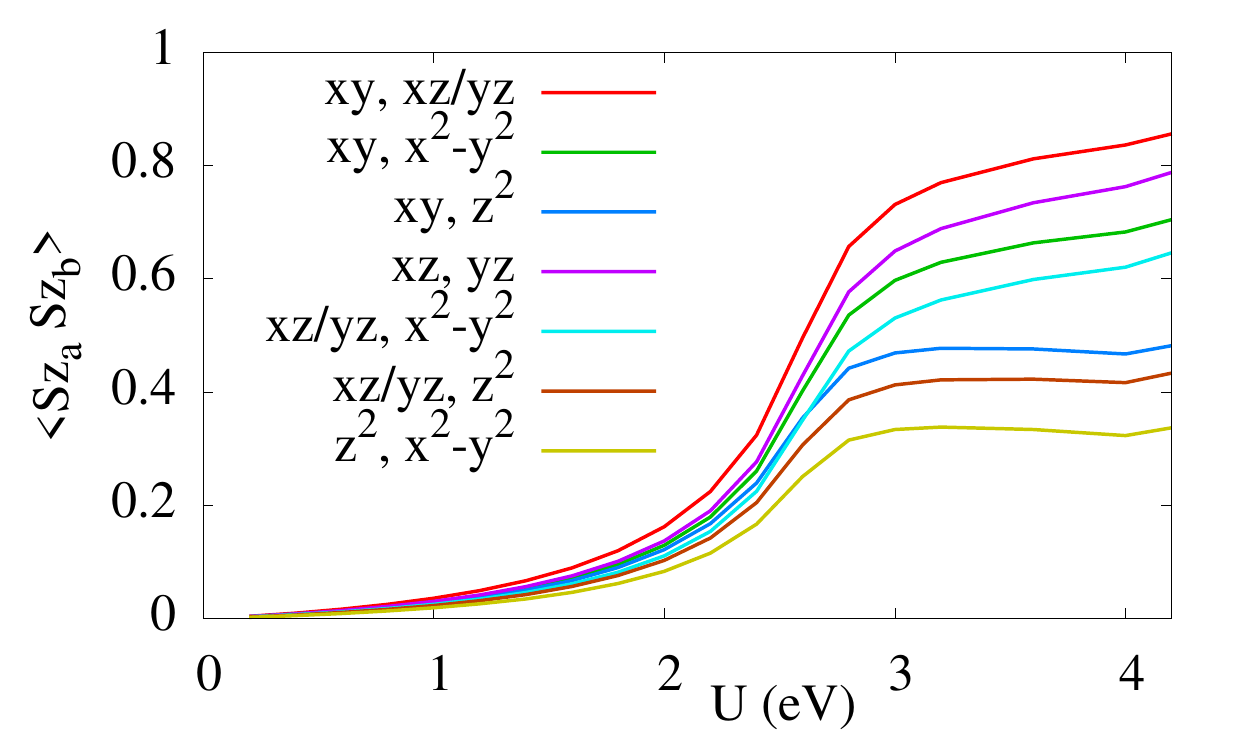}   
  \includegraphics[width=5cm]{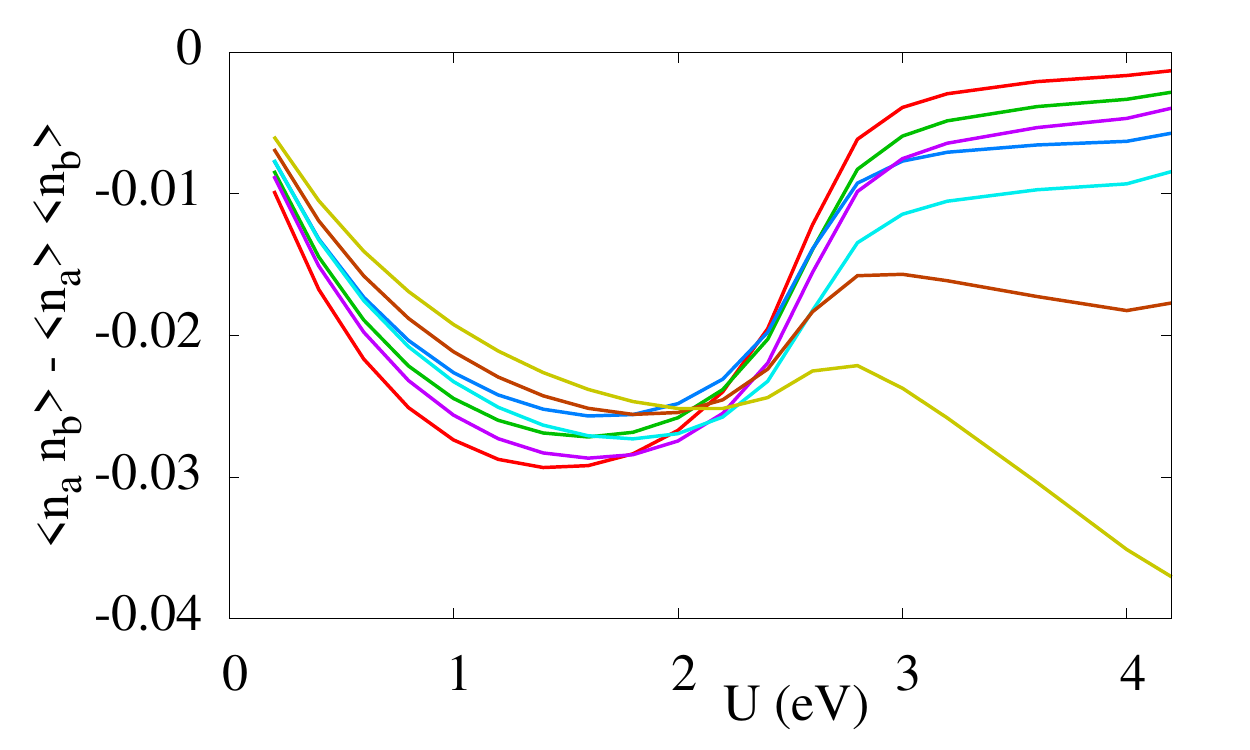}   
    \caption{Orbitally resolved quasiparticle weights and Spin-spin and charge-charge interorbital correlation functions for BaFe${}_2$As${}_2$, as a function of U for J/U=0.25.}
  \label{fig:Z_corr_functs_BaFe2As2}
  \includegraphics[width=5cm]{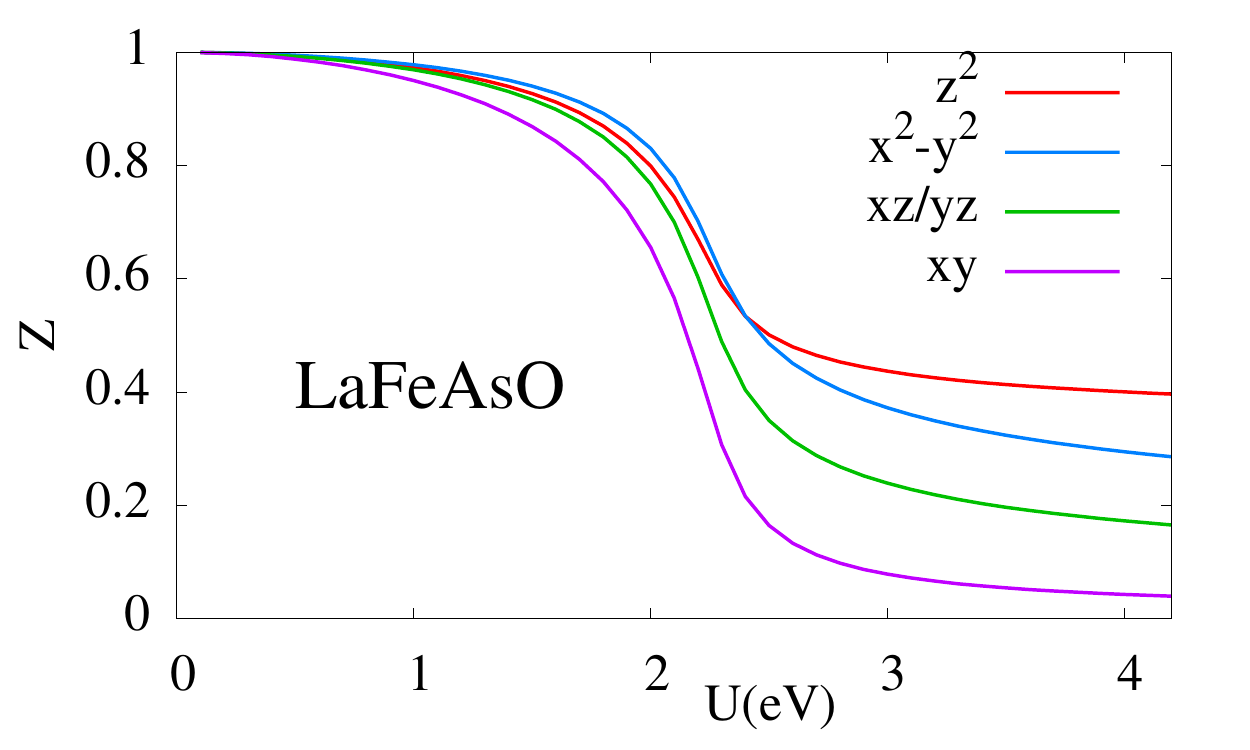}
  \includegraphics[width=5cm]{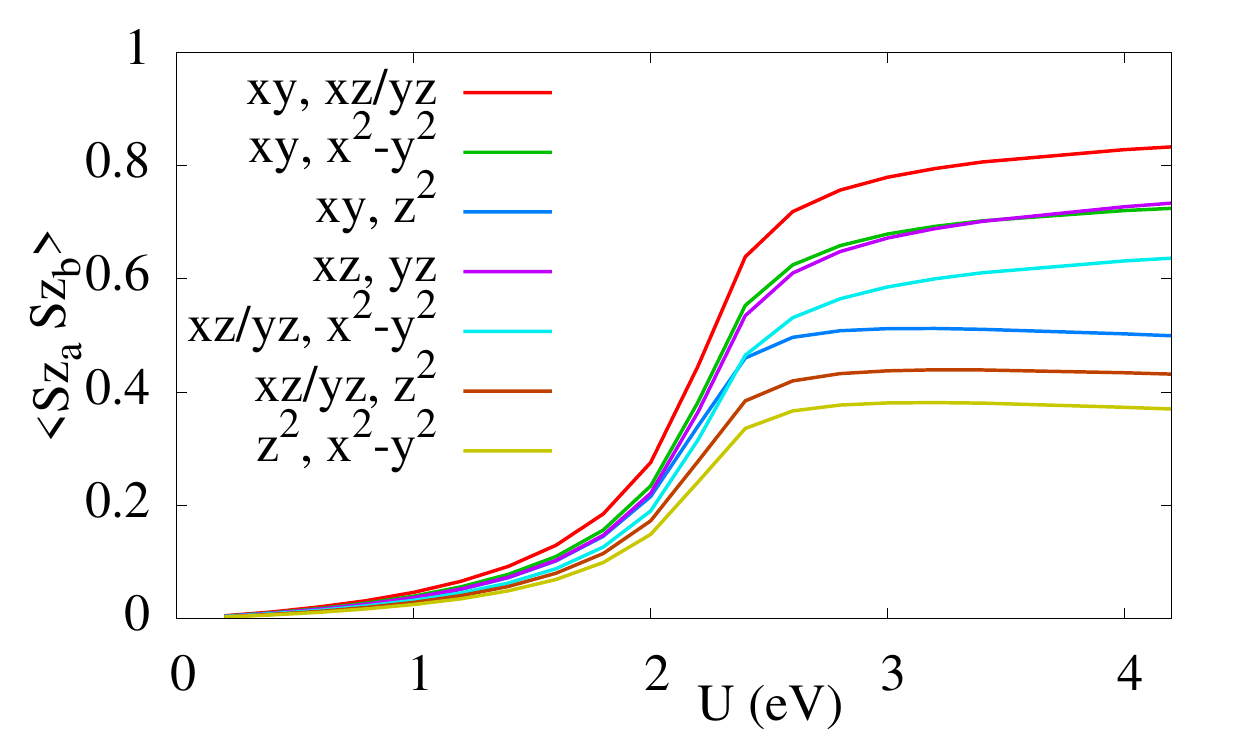}   
  \includegraphics[width=5cm]{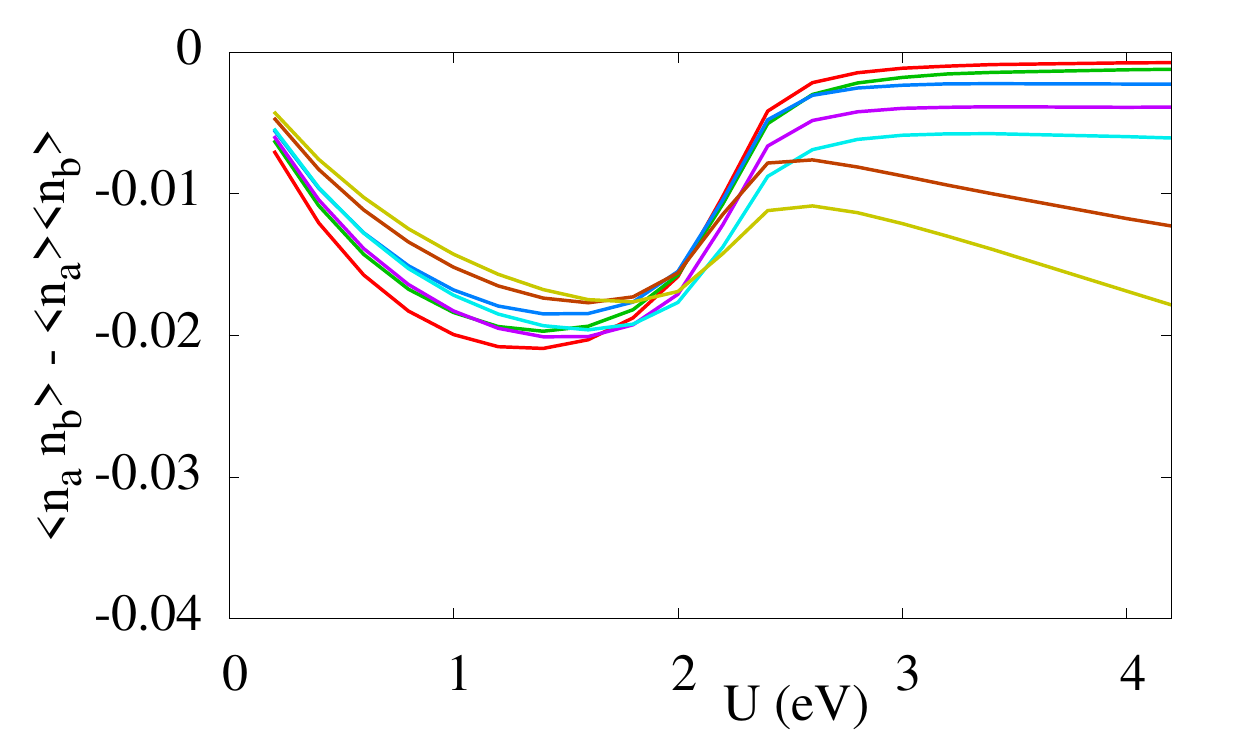}   
  \caption{Orbitally resolved quasiparticle weights and Spin-spin and charge-charge interorbital correlation functions for LaFeAsO, as a function of U for J/U=0.29.}
  \label{fig:Z_corr_functs_LaFeAsO}
  \includegraphics[width=5cm]{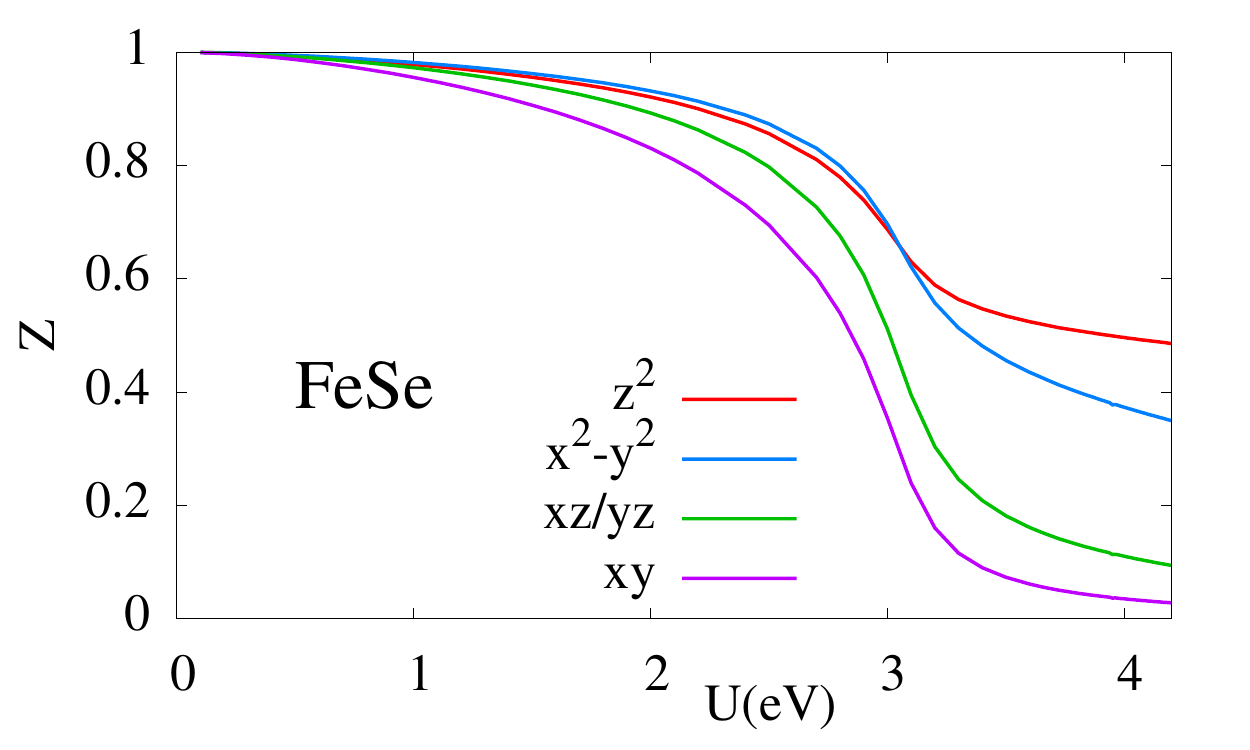}
  \includegraphics[width=5cm]{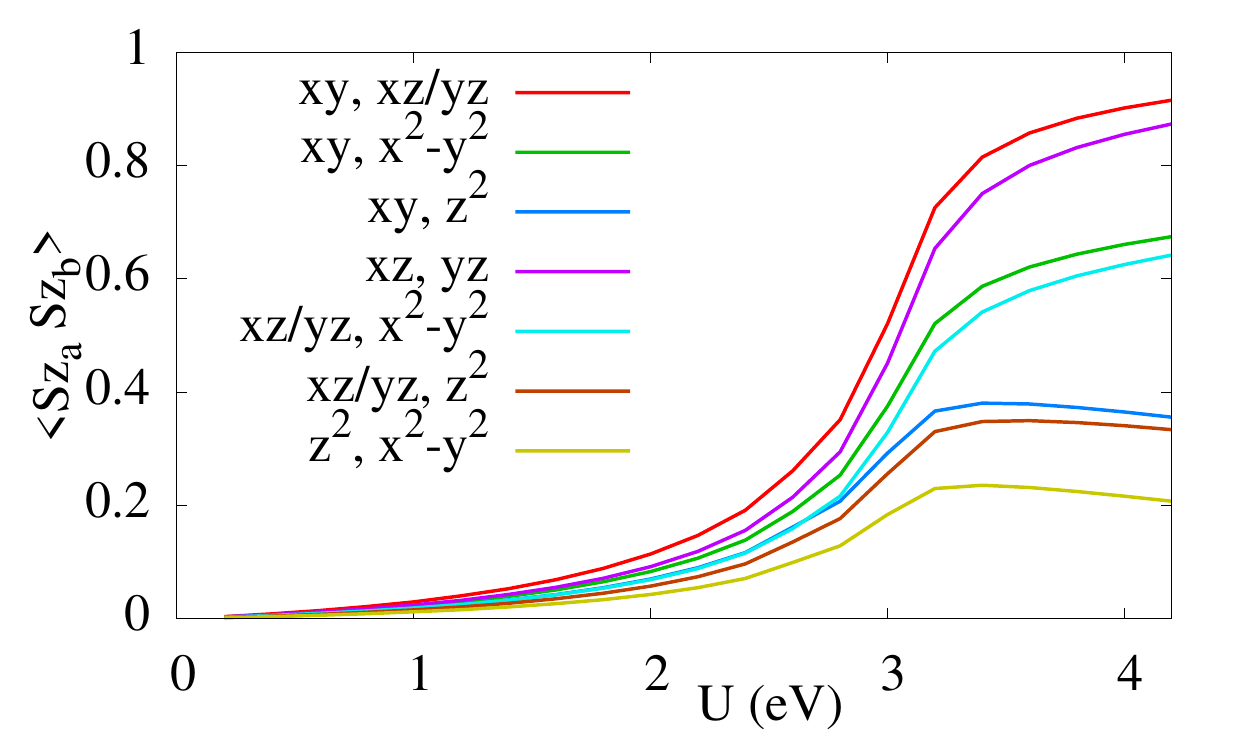}   
  \includegraphics[width=5cm]{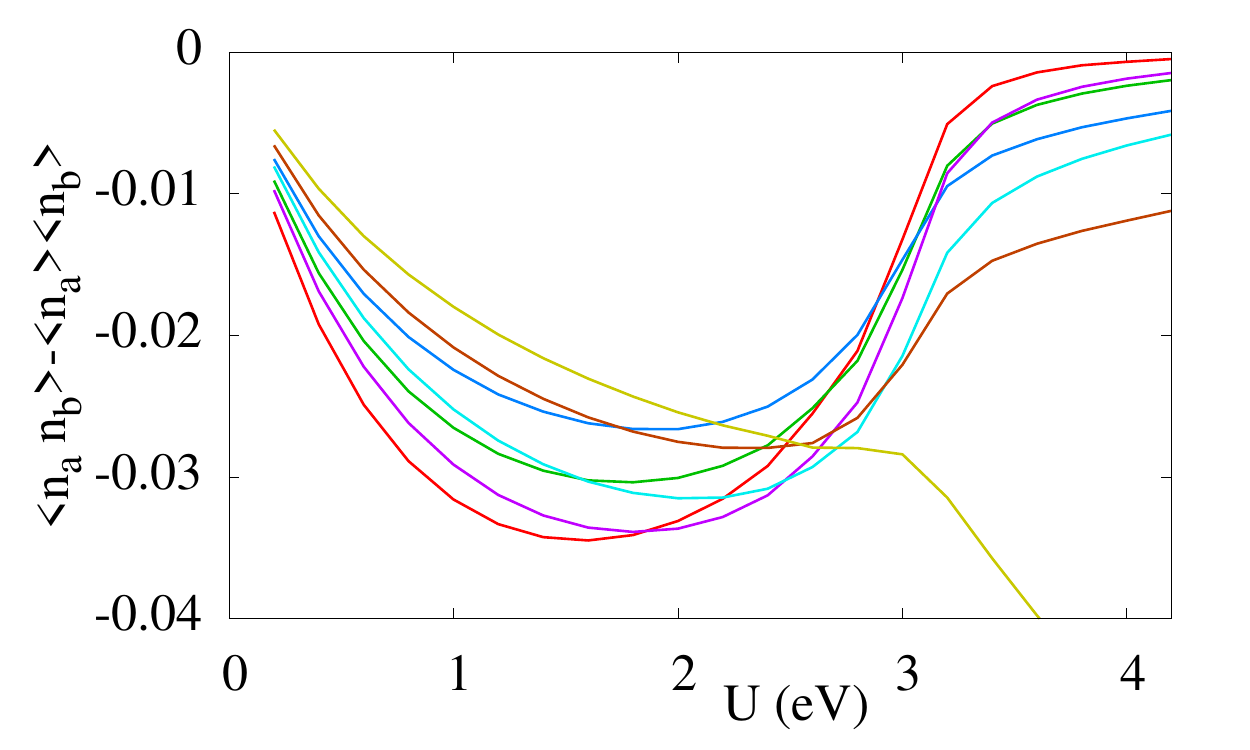}   
    \caption{Orbitally resolved quasiparticle weights and Spin-spin and charge-charge interorbital correlation functions for FeSe, as a function of U for J/U=0.224.}
  \label{fig:Z_corr_functs_FeSe}
\end{center}
\end{figure}
\end{widetext}

\section{Theory}

\subsection{Methods}

The material-specific band structures are determined using Density Functional Theory with 
the Generalized Gradient Approximation for the exchange-correlation potential according to the
Perdew-Burke-Ernzerhof recipe as implemented in Quantum
Espresso~\cite{Giannozzi_QE}. Then we apply Wannier90~\cite{Mostofi_Wannier90} to compute
the maximally localized Wannier orbitals, which we use as a basis to generate a tight-binding Hamiltonian for the 5 orbitals of d-symmetry giving rise to the conduction bands.
We include the electron-electron on-site interaction in the form
\bea\label{H_int}
 H&=&\# U\sum_{i,m}
 n_{im\uparrow} n_{im\downarrow}+U' \@\sum_{i,m>m',\s } n_{im\s} n_{im\bar\s} \\
 &+&(U' -J)\@\sum_{i,m>m',\s } n_{im\s} n_{im'\s} 
\nonumber
 \eea
where $n_{im\s}\equiv d^\+_{im\s}d_{im\s}$ and $d_{im\s}$ is the destruction operator of an electron of spin $\s$ at 
site $i$ in orbital $m$.
$U$ and $U'=U-2J$ are intra- and  inter-orbital repulsions and $J$ is the Hund's coupling, treated here in the Ising (density-density only) form.

This 5-orbital Hubbard-like Hamiltonian is solved within the Slave-Spin mean-field scheme, here generalized to the case of realistic hamiltonians. The original implementation of this method\cite{demedici_Slave-spins,Hassan_CSSMF} (and differently from a more recent implementation\cite{SC_YuSi_LDA-SlaveSpins_LaFeAsO}) yields nonzero Lagrange multipliers in the noninteracting limit, when a crystal-field is present. A constant was added to each Lagrange multiplier in order to compensate this artificial level shift, and adjusted in order to obtain the correct non-interacting population distribution among the orbitals at zero interaction strength.

Details of the implementation will be given elsewhere\cite{GG_DFT+SlaveSpins}.

\subsection{Orbital-decoupling in other iron superconductors}

In this section we compare slave-spin results for stochiometric and doped BaFe${}_2$As${}_2$,  LaFeAsO
and FeSe. 
%Given the approximate nature of our many-body treatment, we did not use ab-initio estimates of the interaction parameters U and J, but we chose them in order to reasonably reproduce the experimental effective mass. In particular we take $U=2.7eV$  for BaFe${}_2$As${}_2$, $U=2.7 eV$ for LaFeAsO \and $U=4.1eV$ for FeSe, which are also close to cRPA estimates for a d-model. We actually first present the results for the stochiometric compounds as a function of U, keeping the ratio fixed at $J/U=0.25$ for BaFe${}_2$As${}_2$, $J/U=0.29$ for LaFeAsO and $J/U=0.224$ for FeSe. 

Given the approximate nature of our many-body treatment (for instance a somewhat larger J is needed in the present formulation of the slave-spin mean-field compared to DMFT) , we
did not use the exact ab-initio estimates of the interaction parameters.
For the Hubbard repulsion we use $U=2.7eV$  for
BaFe${}_2$As${}_2$, $U=2.7 eV$ for LaFeAsO \and $U=4.1eV$ for
FeSe, which are also close to cRPA estimates for a d-model\cite{miyake_interactions_jpsj_2010}, whereas we explore a range of values for the Hund's coupling J. 
For the calculations on BaFe${}_2$As${}_2$ shown in Figure 2 of the main article we have chosen $J/U=0.25$ in order to reasonably reproduce the experimental effective mass at n=6.
We then use $J/U=0.29$ for LaFeAsO  and $J/U=0.224$ for FeSe. 

We first present the results for the stochiometric compounds as a function of U, keeping the
ratio $J/U$ fixed.

These data show a striking similarity between the different compounds:
after an initial slow reduction of the quasiparticle weights Z at weak interaction strength, they all undergo a crossover in which the quasiparticle weights quickly drop and the system enters a different metallic phase (the Hund's metal).
Here the Z are essentially flat and different among the orbitals. The spin-spin correlation functions saturate to high values, indicating the local formation of a (screened) high-spin, and the emergence of a common energy scale for spin-excitations\cite{Greger_OSMT_onescale}, and interorbital charge correlations drop abruptly, indicating the orbital decoupling\cite{SC_demedici_MottHund}.

Some of these results reproduce previous ones \cite{SC_YuSi_LDA-SlaveSpins_LaFeAsO,SC_Lanata_FeSe_LDA+Gutz} and more generally the tendency to orbital differentiation is indeed confirmed by a large amount of realistic simulations performed with several methods for treating the electronic correlations on top of DFT bandstructures: DMFT\cite{SC_Haule_pnictides_NJP,Shorikov_LaFeAsO_OSMT,Laad_SusceptibilityPnictides_OSM,Craco_FeSe,Aichhorn_FeSe,SC_Yin_kinetic_frustration_allFeSC,SC_Ishida_Mott_d5_nFL_Fe-SC, SC_Liebsch_FeSe_spinfreezing},
variational Montecarlo\cite{SC_Misawa_d5-proximity_magnetic},
Slave-spins \cite{SC_YuSi_LDA-SlaveSpins_LaFeAsO,Yu_Si_KFeSe},
Hartree-Fock mean-field\cite{Bascones_OSMT_Gap_halffilling},
fluctuation-exchange approximation\cite{SC_Ikeda_pnictides_FLEX},
Gutzwiller approximation\cite{SC_Lanata_FeSe_LDA+Gutz}.

\begin{figure}[ht]
\begin{center} 
 \includegraphics[width=8cm]{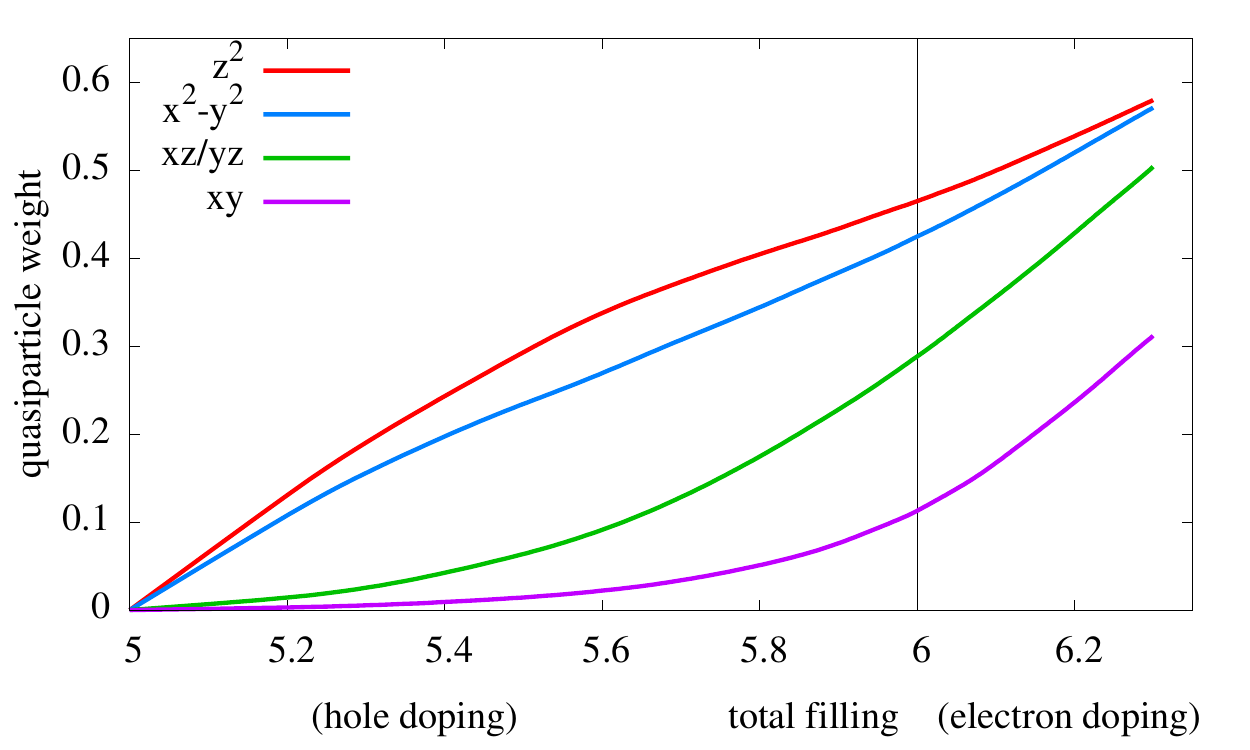}
 \includegraphics[width=8cm]{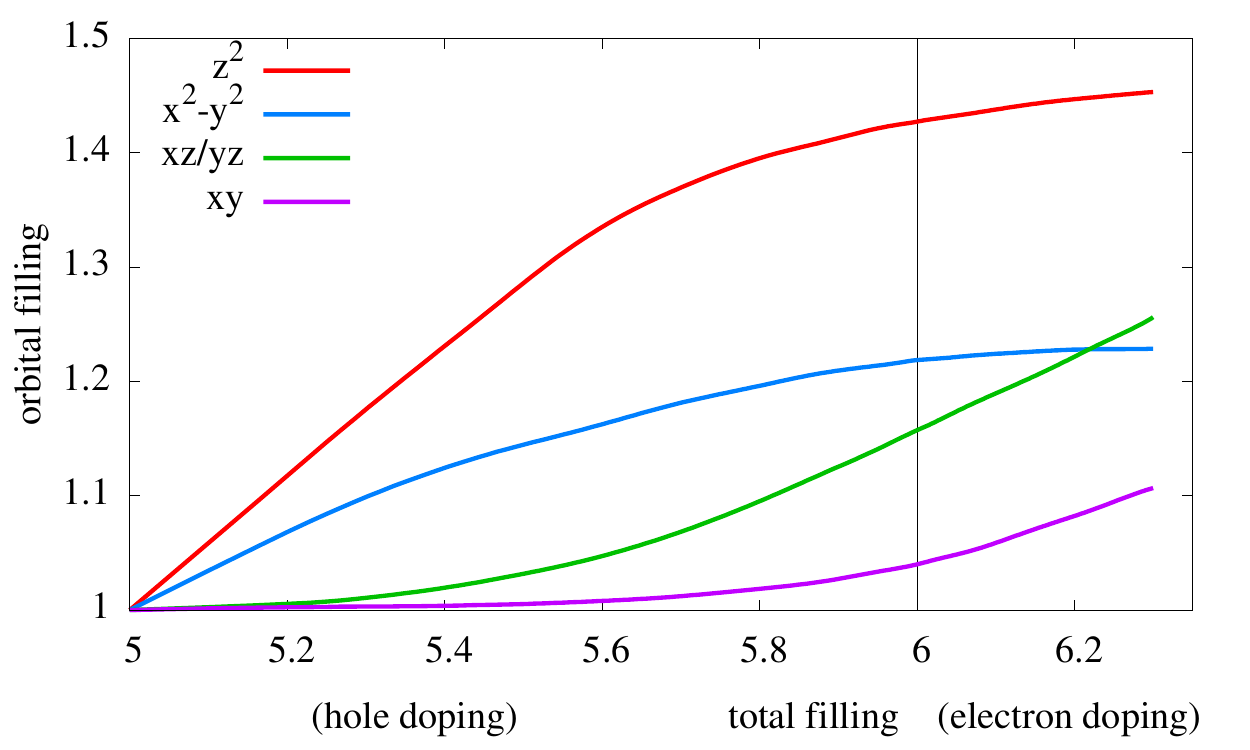}
 \includegraphics[width=8cm]{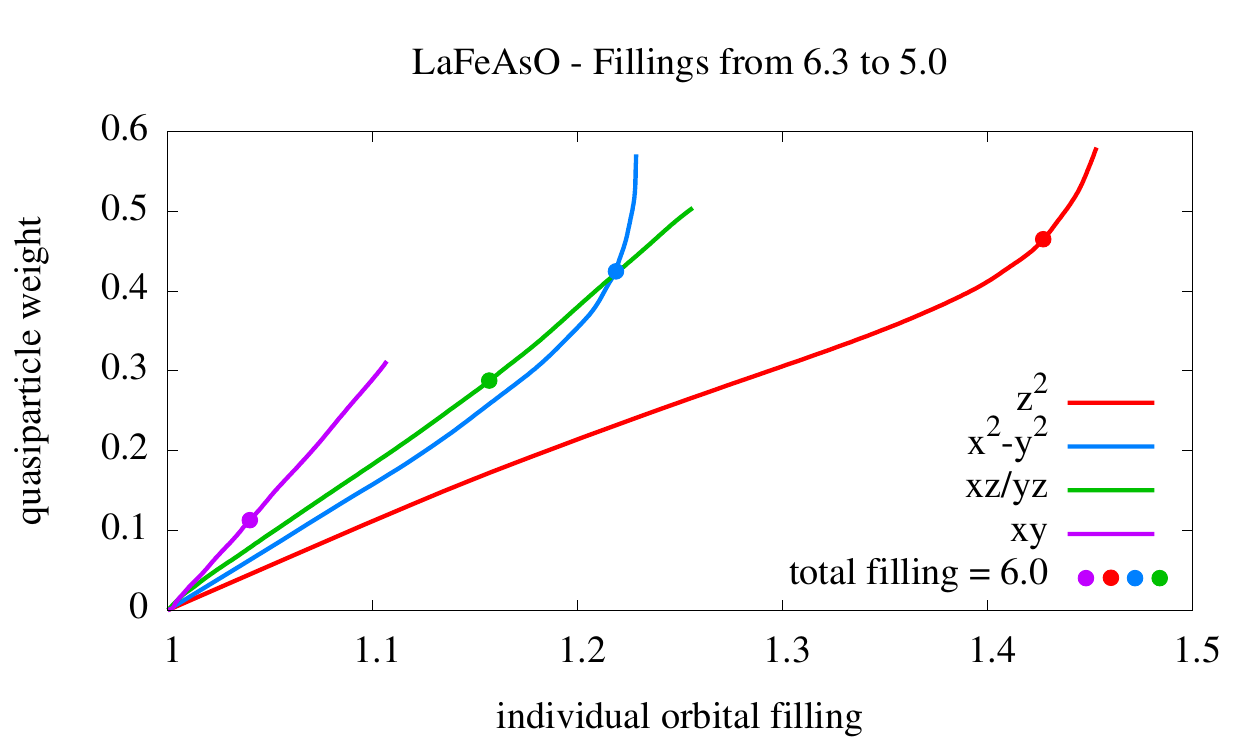}   
    \caption{Quasiparticle weights  $Z_\a$ and orbital populations $n_\a$ (upper panels) as a function of doping for LaFeAsO. $Z_\a$ as a function $n_\a$ for each orbital $\a$ (lower panel). The striking linear behaviour signals the orbital-decoupling.} 
\label{fig:supp_LaFeAsO}
\end{center}
\end{figure}

\begin{figure}[ht]
 \begin{center} 
   \includegraphics[width=8cm]{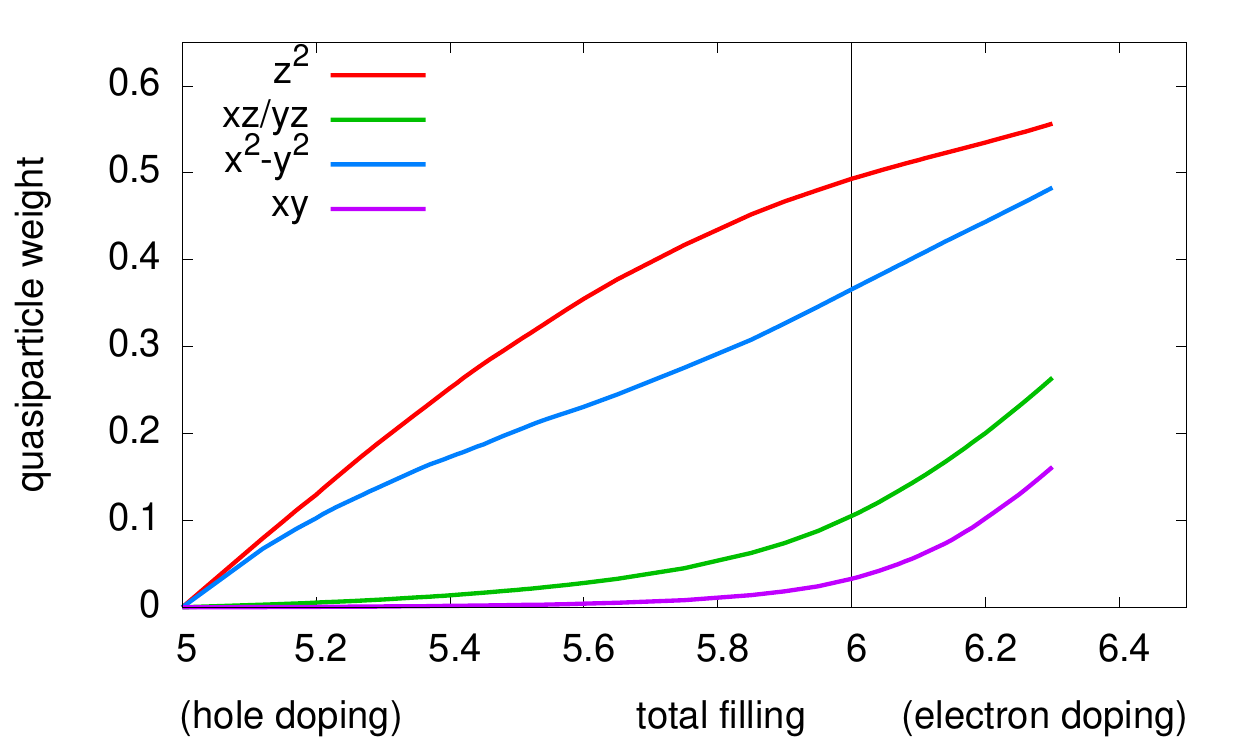}
   \includegraphics[width=8cm]{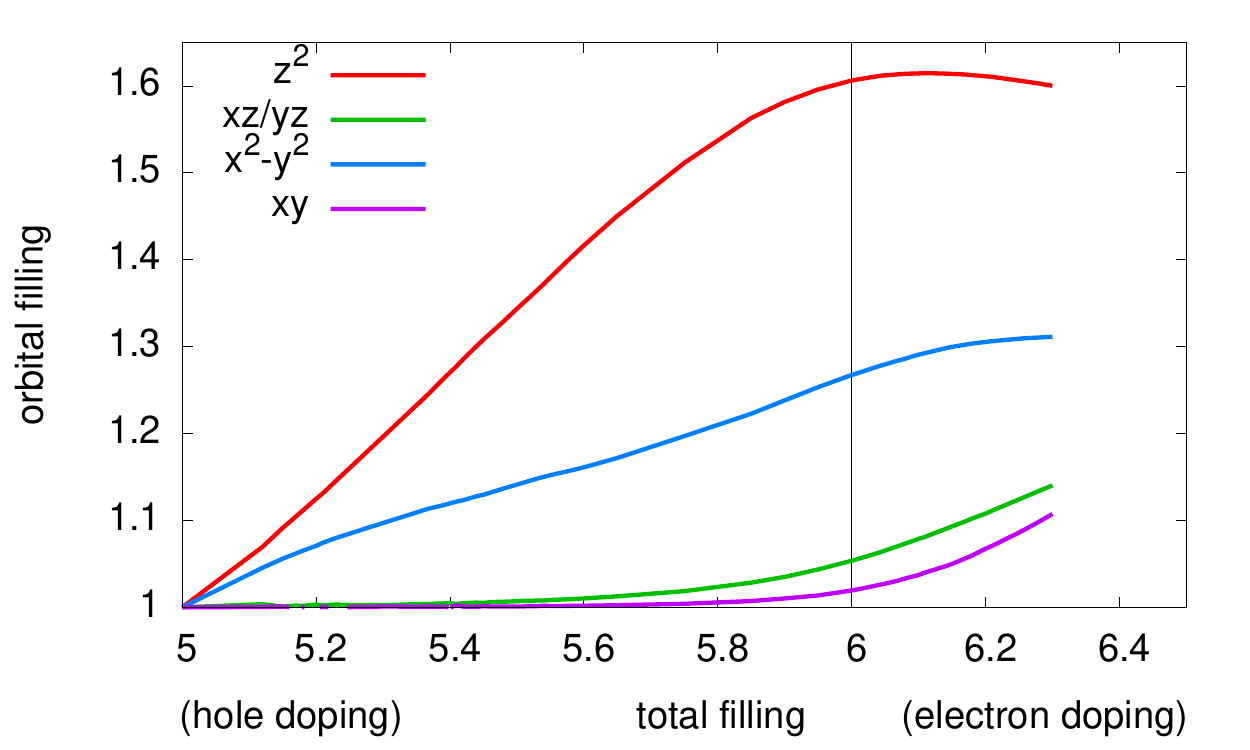}
    \includegraphics[width=8cm]{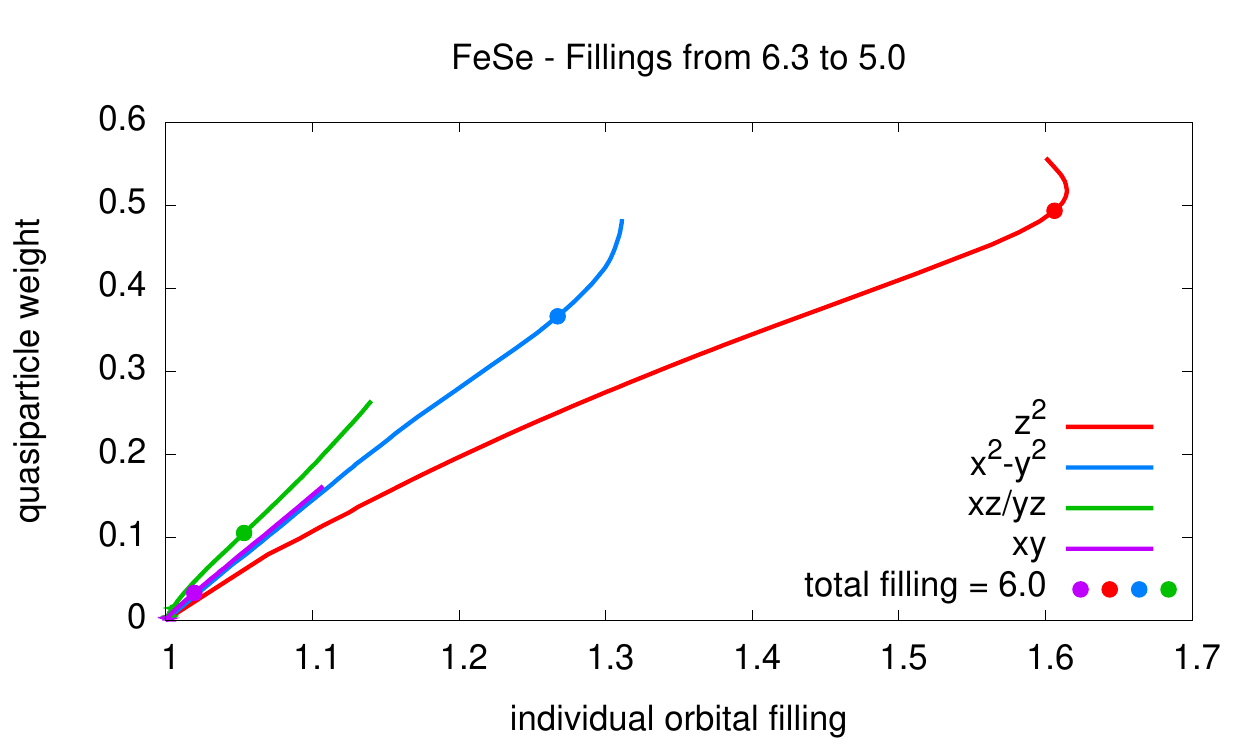}
    \caption{Quasiparticle weights  $Z_\a$ and orbital populations $n_\a$ (upper panels) as a function of doping for FeSe. $Z_\a$ as a function $n_\a$ for each orbital $\a$ (lower panel). The striking linear behaviour signals the orbital-decoupling.} 
 \label{fig:supp_FeSe}
\end{center}
\end{figure}

All the studied stochiometric Iron-superconductors (representative of the `122', `1111' and `11' families) appear to be in the orbitally decoupled bad metal state. 
In order to confirm that the scenario we have drawn in the main
manuscript is general to iron superconductors, we performed calculations as a function of doping for each compound. 
Indeed, as illustrated in Figs \ref{fig:supp_LaFeAsO} and
\ref{fig:supp_FeSe} both LaFeAsO and FeSe show a very similar
behaviour to the one reported for BaFe${}_2$As${}_2$ in Fig.2 of the
main article. The orbitally resolved quasiparticle weights are
linearly dependent on the population of each orbital. Indeed
correlations increase when reducing the total population, the system
approaching the n=5 Mott insulator. Some of these results, such the decreading coherence temperature and orbital fluctuations when approaching half-filling, are already visible in a simplified way in the 3-band Hubbard model\cite{SC_Werner_spinfreezing, SC_demedici_Janus, Yin_PowerLaw}, thus corroborating their generality.

\bigskip

It is also worth pointing out that another, different theory for the differentiation of the correlation degree in iron pnictides, and based on the effect of spin-orbit interaction (that we neglect in this work) was proposed in Ref. \cite{Wu_FeAs_OSMT}.

\subsection{The orbitally-decoupled description of the pseudogap state in cuprates}

A substantial consensus exists around the idea that the basic physics
of cuprate superconductors is that of a doped Mott insulator\cite{Anderson_highTc_RVB},
described by a two-dimensional doped Hubbard model. 

However a crucial feature of this view resides in the importance of short-range correlations. 
Indeed when only local (i.e. uniform in momentum space) correlations are retained - as in the early treatment of the Mott physics in this model (the so-called Brinkman-Rice scenario\cite{Brinkman_Rice}) or its more modern implementations in terms of dynamical mean-field theory\cite{georges_RMP_dmft} - only two paramagnetic phases are found. A Mott insulator is found at half-filling and strong interaction strengths, and a Fermi-liquid elsewhere (in particular at any finite doping and interaction strength), with the effective mass of quasiparticles diverging when approaching the Mott transition (i.e. with a quasiparticle residue characteristically proportional to doping $Z\sim\d$).
When allowing for magnetism an antiferromagnetic long-range order develops at half-filling and at low dopings, but the phase diagram does not show the many other phases found in the cuprates.

A much more realistic scenario is found within theoretical approaches that allow for differentiation of electronic correlations in momentum space. 
It is not within the scope of this article to review all the literature concerning these approaches, ranging from the slave-bosons based treatments of d-wave superconductivity\cite{kotliar_liu_RVB} to cluster extensions of the dynamical mean-field approach \cite{Maier_RMP_Cluster} capable of describing the pseudogap phase at low hole doping.
We will limit ourselves to notice that the treatment of short-ranged
correlations by these approaches is enough to reproduce all the phases
occurring in the phase diagram of cuprates: an antiferromagnetic
(Mott) insulating phase around half-filling, a pseudogapped metallic
phase at low doping with strong k-space anisotropy, a d-wave
superconducting dome at intermediate doping and a rather good
Fermi-liquid phase at high doping (see for example \cite{Gull_Energetics}).
%. These calculations seem therefore to capture all the basic physics
%of high-Tc superconductors, even (see for example Refs. \cite{Gull_DCA_SC_PG, Gull_Energetics}).

The point we make here is that the comparison between these calculations and our present results shows that "orbital-decoupling" is the mechanism behind the formation of the pseudogap and thus that it is a basic common feature of both cuprates and pnictides, albeit showing up in a subtly different way in the two cases.

We base our present discussion on the recent work from Gull et al. \cite{SC_Gull_DCA_k-space_selec} (which builds upon previous work, referenced therein), that treats the normal phase of cuprates by means of a study of the bidimensional Hubbard model as a function of doping within the so-called Dynamical Cluster Approximation, discarding all long-range orders.

Within this method local and short ranged correlations due to the local electronic repulsion are incorporated in the non-interacting low-energy band structure (reproducing the Fermi surface evolution with doping) through a set of self-energy functions, one for each of the N sectors (zones of equivalent k-space volume) in a partitioning of the Brillouin zone. These functions are calculated numerically within an auxiliary self-consistent cluster impurity model of N sites, that can be equivalently formulated as an N-orbital impurity model, in which at every orbital is associated a sector in the Brillouin zone. This method becomes exact for $N\rightarrow \infty$, but it is obviously limited to rather small N (typically $\sim 4\div 16$), for computational reasons. The results quoted here are for the 8-site cluster, and are deemed semi-quantitatively valid for larger N.

Without entering further into the technical details the main results can be summarized as follows. At large doping the self-energies are very close to one another and relatively small at all frequencies, indicating a moderate impact of correlations, uniform at all k vectors. Upon reduction of doping a differentiation of these functions shows larger correlation effects in the sector around $\vk=(\pi,0)$ (and symmetric) than elsewhere. Here (at the "antinodal regions" or "hot spots") quasiparticles at the Fermi surface have a reduced lifetime compared to the others. At still smaller doping a (pseudo-) gap to electronic excitations opens in this sector (through a divergence of the corresponding self-energy at small frequencies), whereas the other sectors remain comparatively more weakly correlated. In particular the sector around $\vk=(\pi/2,\pi/2)$ where the rest of the Fermi surface lies (the "nodal regions" or "cold spots") still has relatively well-defined quasi-particles. Finally, at half-filling this last sectors open a gap too, and the system becomes an insulator.

It has already been noticed that this description of the pseudogap phase is very much analogous to an "Orbital-selective Mott transition"\cite{Biermann_nfl, Ferrero_dimer}, found in multi-orbital systems when a gap opens for excitations in a subset of orbitals, while quasiparticles still exist in the rest of the system. This similarity extends then to iron superconductors\cite{SC_Ishida_Mott_d5_nFL_Fe-SC}, in particular supported by the theoretical and experimental scenario of selective Mottness, i.e. of very strong differentiation of correlations - even if a strict OSMT does not occur -, presented in this paper.

What we highlight here is that this analogy is clearly deeper than one might suspect, and that orbital-decoupling is the mechanism behind both phenomenons.

\begin{figure}[ht]
 \begin{center} 
   \includegraphics[width=8cm]{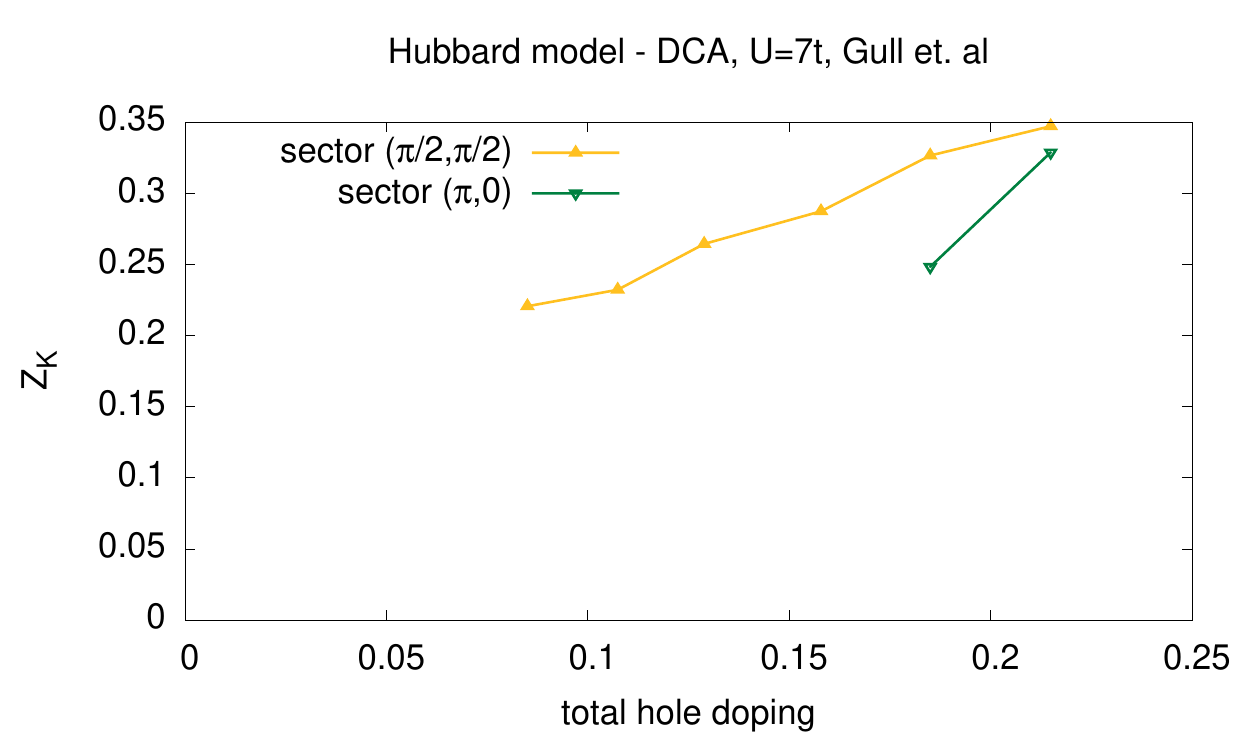}   
   \includegraphics[width=8cm]{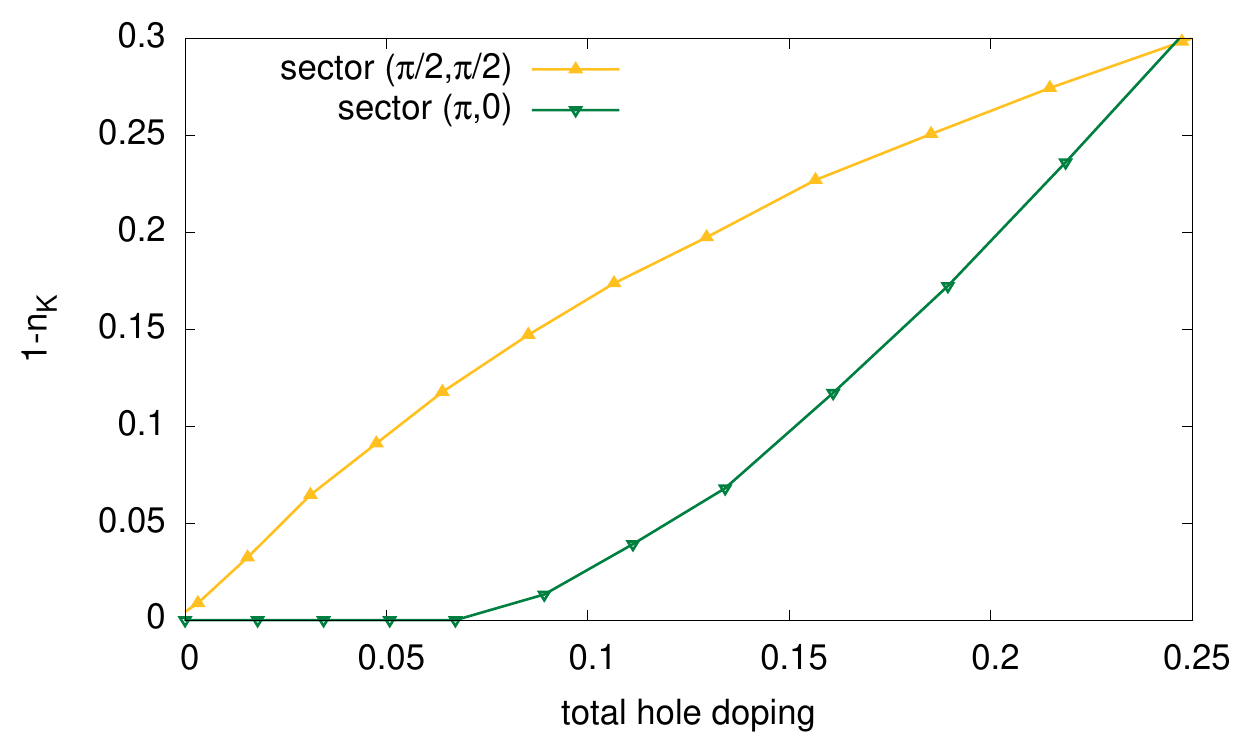}  
    \includegraphics[width=8cm]{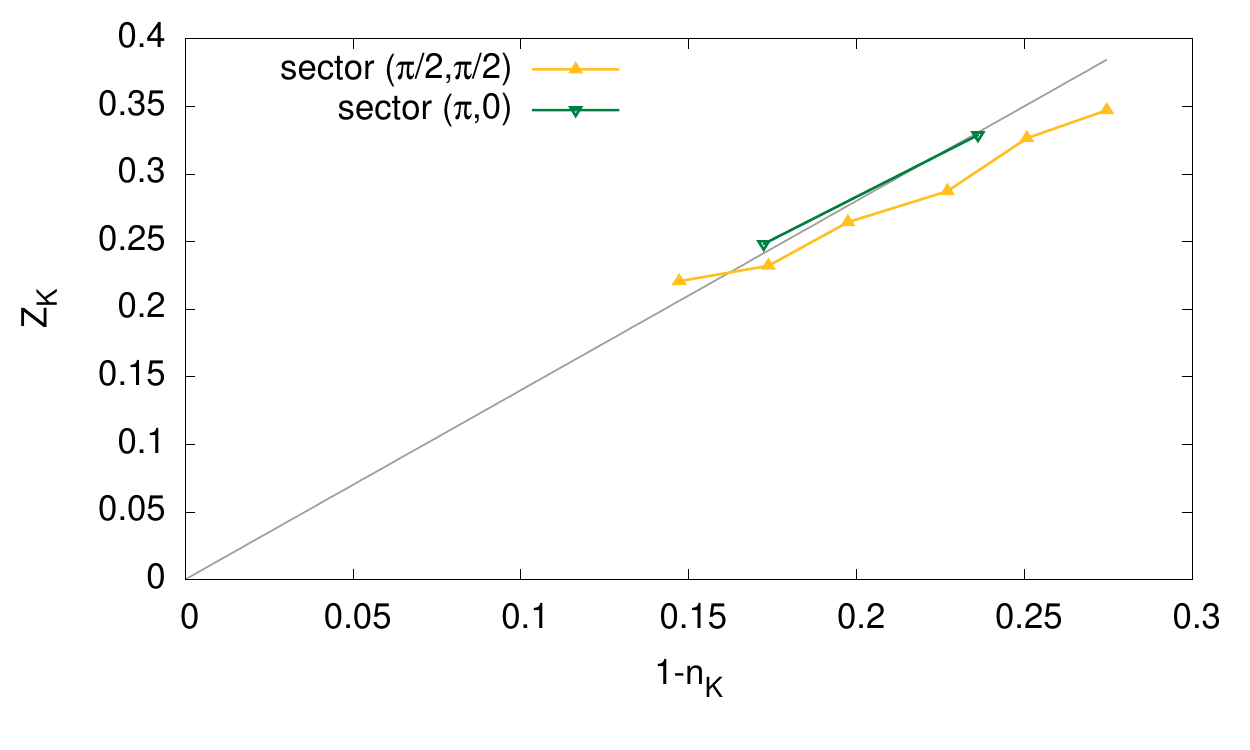}
    \caption{DCA calculations for the single-band Hubbard model for U=7t, from Gull et al.\cite{SC_Gull_DCA_k-space_selec} Weight (and inverse mass enhancement)  $Z_K$ for the quasiparticles in the sector K of the Brillouin zone (upper panel) and corresponding normalized hole doping $1-n_K$ (see text, middle panel). Lower panel: quasiparticle weight plotted there as a function of the corresponding $1-n_K$. Within the limited published data we observe a linear behaviour signaling the orbital-decoupling. The solid line is a guide to the eye.} 
 \label{fig:Z_DCA}
\end{center}
\end{figure}
Indeed we report in Fig. \ref{fig:Z_DCA} the data from Gull et al.\cite{SC_Gull_DCA_k-space_selec}, and in particular we plot the weight $Z_K$ for the quasiparticles in the antinodal $(\pi,0)$ and nodal $(\pi/2,\pi/2)$ sectors, as a function of hole doping.  Both sectors will develop a gap at half-filling, and even if a divergence of the effective mass is not expected due to the effect of short-range correlations, in the range of doping where Gull et al. could trace $Z_K$ (the finite temperature in their calculations limits this to doping values where $Z_K$ is not too low), for both sectors the quasiparticle weight decreases with decreasing doping. Around doping $\sim 0.07$ the antinodal sector becomes gapped, and the system enters the sector-selective Mott state. This is also signaled by the corresponding orbital in the impurity model becoming half-filled, with population (the population for each sector is normalized, so that the total population of the system is obtained by the formula $n=1/N\sum_K n_K$) insensitive to the chemical potential, thus indicating a (sectorwise) incompressible state (middle panel in Fig. \ref{fig:Z_DCA}).

This scaling of the quasiparticle spectral weight with the population of the sector/orbital approaching half-filling is clearly reminiscent of the orbital-decoupling mechanism outlined in this paper.
In order to check this hypothesis we plot, as we did for iron superconductor calculations, the quasiparticle weight in each sector as a function of its individual doping from half-filling. Here too we obtain (lower panel of Fig. \ref{fig:Z_DCA}) a quite linear behavior for $Z_K(n_K)$ with a very small or vanishing intercept, showing that each sector follows the conventional dependence of a doped Mott insulator. 
The possible (in particular for the nodal sector) small deviations from the theoretical $Z_K \propto n_K$ seem to signal departures from a perfect decoupling. It is however to be signaled that the value of $U=7t$ chosen (for computational reasons) in the work of Gull et al. is smaller than the more realistically expected $U=9t$, that should then enhance the decoupling.
Also, for very small doping a departure from the linear behavior of $Z_K$ is ultimately expected in order to avoid the divergence of the effective masses.

This result is reported in the lower panel of Fig. 2 in the main text, and compared very favourably with the one for pnictides. 

Thus we conclude that the DCA treatment of the single band Hubbard model, that seems to capture all the most important features of the physics of cuprates, shares the striking aspect that we highlight in this paper for the physics of iron superconductors: the orbitals of the auxiliary model, corresponding each one to a different sector of the Brillouin zone are essentially 'decoupled' from one another, in the same way as the physical orbitals are in iron superconductors. As a result their degree of correlations is regulated by their individual distance from half-filling.

Thus cuprates and iron superconductors, seem to share a very basic mechanism, essential for the properties of the respective normal phases, and possibly relevant for the superconductive ones.
From this viewpoint the phase diagrams of the two compounds, once anlyzed as we do in Fig. 3 of the main text, share more than a superficial analogy, and their similarity in the light of the orbital-decoupling mechanism are hardly a coincidence.

\end{document}